\documentclass[final,3p,sort&compress,times]{elsarticle}
\usepackage{lineno,hyperref}
\usepackage{graphicx}% Include figure files
\usepackage{dcolumn}% Align table columns on decimal point
\usepackage{bm}% bold math
\usepackage{upgreek}
\usepackage{amsmath}
\usepackage{multirow}
\usepackage{multicol} % Multiple columns environment

\usepackage{nomencl} % Nomenclature package
\usepackage{hyperref}

\usepackage{etoolbox}
\renewcommand\nomgroup[1]{%
  \item[\bfseries
  \ifstrequal{#1}{G}{Greek symbols}{%
  \ifstrequal{#1}{S}{Subscripts}{}}%
]}

\newcommand{\Nu}{N\!u}%
\journal{International Journal of Heat and Mass Transfer}

\begin{document}

\begin{frontmatter}

\title{Interaction between vapor bubbles during flow boiling heat transfer in microchannels}

\author[SKLE]{Odumuyiwa A. Odumosu}
\author[SKLE]{Mengqi Ye}
\author[SKLE,NIEPES]{Tianyou Wang}
\author[SKLE,NIEPES]{Zhizhao Che\corref{cor1}}
\cortext[cor1]{Corresponding author.}
\ead{chezhizhao@tju.edu.cn}
\address[SKLE]{State Key Laboratory of Engines, Tianjin University, Tianjin, 300350, China.}
\address[NIEPES]{National Industry-Education Platform of Energy Storage, Tianjin University, Tianjin, 300350, China}

\begin{abstract}
Microchannel flow boiling is an efficient cooling solution for high-power-density miniaturized systems. Many studies on microchannel flow boiling focused on the dynamics of single vapor bubbles, while neglecting the interaction between bubbles, which is important in relevant applications. Here, numerical simulations are carried out to study the interaction between multiple vapor bubbles in microchannel flow boiling. The results show that for different numbers of bubbles in the microchannels with the same initial size and position of leading bubbles, the bubble size in a single-bubble microchannel is larger compared to the leading bubble of multiple-bubble cases because of heat absorption by the vaporization at the rear bubbles. As the initial volume ratio between the leading bubble and the rear bubble decreases, the leading bubble size in the downstream becomes smaller because of the reduced contact with the superheated thermal boundary layer. With increasing the Reynolds number, both the leading and the trailing bubbles increase slightly in size in the upstream of the heated region, because the bubbles at higher Reynolds number move faster and firstly get in contact with the superheated fluid. The increase in the bottom wall thickness increases the growth rate of the multiple bubble sizes with earlier bubble coalescence because of the higher upstream wall temperature by heat conduction in the solid wall.
\end{abstract}

\begin{keyword}
\texttt {Bubble interaction \sep Boiling heat transfer \sep Bubble growth \sep Microchannel \sep Conjugate heat transfer 
}
\end{keyword}
\end{frontmatter}

\section{Introduction}\label{sec:1}
The removal of high heat flux emanating from numerous miniaturized high-power-density devices during operation poses huge thermal risks if not managed optimally. The challenging task arises due to the dissipation of heat from confined small areas. Microchannel flow boiling offers a cutting-edge solution to dissipate the high heat flux effectively at a microscale area by utilizing the working fluid latent heat and the high surface-to-volume ratio \cite{odumosu23, tang22}. The intricacy between heat transfer, fluid dynamics, and phase change in microchannel flow boiling requires a thorough understanding of this promising yet complex area of research \cite{zhang24}. Hence, numerous studies have been performed to analyze the microchannel flow boiling processes \cite{baldassari13, Bertsch2008ReviewFlowBoiling, garimella03, guo14, harirchian11, kadam21, nahar21, thome04}.

Although many experiments have been conducted to study the flow boiling in microchannels \cite{Bigham2015FlowBoiling, Gedupudi2011BubbleGrowth, halon22, kumar22, rui23, zhang20, zhang17, zhou21}, the main drawback to experiments is the limitation of high-resolution instruments essential to describe the flow boiling behavior at small temporal and spatial scales, coupled with extensive multi-physics and complex processes \cite{luo20}. In contrast, numerical methods via the computational fluid dynamics (CFD) analysis can access complex details of the flow boiling in microchannels. Mukherjee et al.\ \cite{mukherjee11} studied numerically the microchannel flow boiling of a vapor bubble growth and found that the motion of the liquid-vapor interface evaporation enhanced the heat transfer. Zhuan and Wang \cite{Zhuan2012FlowBoilingMicrochannel} investigated the transitions of flow regimes during flow boiling in circular microchannels and observed that bubble growth and coalescence are significant factors for flow transitions. Ferrari et al.\ \cite{Ferrari2018SlugFlowBoiling} studied numerically the influence of the channel geometry during flow boiling of R245fa and showed that the channel geometry significantly impacts the heat transfer and the bubble dynamics. Magnini and Matar \cite{magnini20} studied numerically the effect of aspect ratio on microchannel flow boiling and found that as the aspect ratio increases, the square microchannels at low capillary numbers have very thin liquid films between the bubble and the channel wall than when compared to rectangular microchannels which have thicker liquid films. Magnini et al.\ \cite{Magnini2013ElongatedBubbleFlowBoiling} implemented the height function algorithm to reconstruct the interface in the simulation of microchannel flow boiling in slug flow regime and found that the main mechanism of heat transfer with elongated bubbles is thin-film evaporation. Guo et al.\ \cite{guo16} conducted numerical modeling of microchannel flow boiling in annular flow regime results and concluded that their model agreed with the film thickness correlations in the literature. Luo et al.\ \cite{luo20} considered annular flow boiling, and found the liquid film between the interface and the wall becomes thinner as the wall heat flux or inlet vapor quality increases. More recently, Priy et al.\ \cite{priy24} performed 2D numerical investigations on microchannel flow boiling to analyze the bubble nucleation and interaction involving single and multiple nucleate sites. Their numerical model captured the microchannel flow boiling processes including the vapor bubble generation, growth, departure, coalescence, and sliding within the channel. Zhang et al.\ \cite{zhang24} investigated the intricacy between heat transfer and flow dynamics of boiling in curved microchannels. They found that curved channels enhance heat transfer compared to straight channels.

Despite numerous studies on microchannel flow boiling, most of the previous studies have been devoted to the behavior of single vapor bubbles. The interaction between vapor bubbles in real applications is important because it affects not only the bubble dynamics, but also the phase change and the boiling heat transfer efficiency \cite{widyatama23}. The interaction between two bubbles in a vertical microchannel was studied numerically by Liu and Palm \cite{liu16}. They used the coupled level set and volume of fluid (CLSVOF) method for interface capturing, and found that the bubble dynamics and heat transfer are influenced by the interacting processes involving the sliding, merging, and post-merging stages. Consolini and Thome \cite{consolini10} developed a numerical model to study the flow boiling heat transfer of coalescing bubbles in a microchannel. Their model predicted the heat transfer mechanisms and the bubble growth. Magnini et al.\ \cite{magnini13} conducted a numerical study to investigate the influence of leading and sequential bubbles in microchannel flow boiling. They found that the presence of multiple bubbles enhanced the heat transfer and described the complex dynamics of the slug flow boiling regime. Liu et al. \cite{Liu2017BubbleTrainMicrochannel} simulated the dynamics of the bubble train in microchannel flow boiling and found that the bubbles’ growth was non-uniform while moving downstream due to different contact positions of the bubbles with the thermal boundary layer. Lombaard et al.\ \cite{lombaard21} studied numerically the interaction between multiple vapor bubbles and found that the heat transfer was enhanced with sequential multiple vapor bubbles present in the microchannel. 

Even though some efforts have been made to understand the interaction between bubbles in microchannel boiling heat transfer. The detailed mechanism is still not clear, and the effect of some key factors is still unknown. The interaction of bubbles in flow boiling is a complex process because of the phase change of the working fluid, the rapid evolution of the interface, and the conjugate heat transfer of conduction in the solids and convection in the fluids. Therefore, this study focuses on the interaction between multiple vapor bubbles in microchannel flow boiling with conjugated heat transfer. Further, the effects of the bubble volume ratio, the inlet Reynolds number, and the bottom wall thickness on the interaction of vapor bubbles and heat transfer are analyzed. This study emphasizes the remarkable effect of the interaction of bubbles on the flow dynamics and thermal performance of microchannel heat transfer. The interaction of multiple bubbles can lead to coalescence and change in the flow patterns, which is potential to affect the thermal performance of microchannel heat sinks. The understanding of bubble interactions underscores the importance of multiple bubbles flow dynamics which behave differently compared to isolated bubbles.

\section{Numerical method}\label{sec:2}
\subsection{Numerical model}\label{sec:2.1}
In the present study, OpenFOAM is utilized to perform the simulation of the interaction between vapor bubbles during flow boiling heat transfer in microchannels. The multi-region solver (multiRegionPhaseChangeFlow) developed by Scheufler and Roenby \cite{scheufler23} is adopted for the simulation. The solver calculates the conservation equations of the mass, momentum, energy, and phase fraction with phase change phenomena and conjugated heat transfer. The numerical model comprises the simultaneous solution of both the fluid and solid regions of the computational domain. In the fluid region, the conservation equations of the mass, momentum, and energy are solved, while only the energy equation is solved within the solid region. The mass and momentum conservation equations for the fluid region are
\begin{equation}\label{eq:01}
\frac{\partial (\rho )}{\partial t}+\nabla \cdot (\rho \mathbf{u})=0
\end{equation}
\begin{equation}\label{eq:02}
\frac{\partial (\rho \mathbf{u})}{\partial t}+\nabla \cdot (\rho \mathbf{uu})=-\nabla p-\nabla \cdot \{{{\mu }_{\text{eff}}}[\nabla \mathbf{u}+{{(\nabla \mathbf{u})}^{T}}]\}+\rho \mathbf{g}+{{\mathbf{f}}_{\sigma }}
\end{equation}
where ${{\mu }_{\text{eff}}}$ is the fluid viscosity, and ${{\mathbf{f}}_{\sigma }}$ is the surface tension force. The energy conservation equation adopts a two-field approach in terms of the temperature,
\begin{equation}\label{eq:03}
\frac{\partial (\rho {{c}_{p}}T)}{\partial t}+\nabla \cdot (\rho {{c}_{p}}\mathbf{u}T)=\nabla \cdot (k\nabla T)-{{\dot{q}}_{pc}}
\end{equation}
where $k$ is the thermal conductivity, ${{c}_{p}}$ is the specific heat capacity, and ${{\dot{q}}_{pc}}$ is for energy changes caused by phase change. 

The volume of fluid (VOF) method is used to predict the evolution of the liquid-vapor interface,
\begin{equation}\label{eq:04}
\frac{\partial \alpha }{\partial t}+\nabla \cdot (\mathbf{u}\alpha )={{\dot{\alpha }}_{pc}}
\end{equation}
where $\alpha$ is the volume fraction of the liquid phase, and ${{\dot{\alpha }}_{pc}}$ is the explicit source term to account for phase change. 

The thermophysical parameters of the fluid are calculated from the volume fraction 
\begin{equation}\label{eq:05}
\varphi =(1-\alpha ){{\varphi }_{v}}+\alpha {{\varphi }_{l}}
\end{equation}
where $\varphi $ represents any fluid properties, including viscosity, density, and thermal conductivity.

The surface tension effect is modeled by including a source term ${{\mathbf{f}}_{\sigma }}$ in Eq. (2) at the liquid-vapor interface \cite{Brackbill1992ModelingSurfaceTension}
\begin{equation}\label{eq:06}
{{\mathbf{f}}_{\sigma }}=\sigma \kappa \mathbf{n}\left| \nabla \alpha  \right|\frac{2\rho }{{{\rho }_{v}}+{{\rho }_{l}}}
\end{equation}
where $\kappa $ and $\mathbf{n}$ are the curvature and the unit normal vector of the liquid-vapor interface. 
\begin{equation}\label{eq:07}
\kappa =-\nabla \cdot \mathbf{n}
\end{equation}
\begin{equation}\label{eq:08}
\mathbf{n}=\frac{\nabla \alpha }{\left| \nabla \alpha  \right|}
\end{equation}

The Hardt and Wandra model \cite{hardt08} was used to calculate the mass transfer due to phase change at the liquid-vapor interface and the VOF equation to account for the liquid loss, and the details can be found in Refs.\ \cite{hardt08, scheufler23}. 
The phase change model calculates the mass transfer at the liquid-vapor interface as a one-species system, which provides the source terms for the continuity, energy, and volume of fluid equations. The energy source term in Eq.\ (\ref{eq:03}) is computed implicitly by the gradient-based model approach that improves the solver’s stability and is expressed as
\begin{equation}\label{eq:09a}
\dot{q}_{pc}=q_{pc}^{l}+q_{pc}^{v}={{k}^{l}}\nabla {{T}^{l}}\cdot {{\mathbf{\hat{n}}}_{s}}+{{k}^{v}}\nabla {{T}^{v}}\cdot (-{\mathbf{\hat{n}}_{s}})
\end{equation}
where ${\mathbf{\hat{n}}_{s}}$ is the interface normal. The mass transfer in Eq.\ (\ref{eq:04}) is evaluated as 
\begin{equation}\label{eq:10a}
{{\dot{\alpha }}_{pc}}=\frac{{\dot{\rho }}}{{{\rho }^{l}}}
\end{equation}
\begin{equation}\label{eq:11a}
\dot{\rho }=\frac{{{q}_{pc}}\left| {{n}_{s}} \right|}{{{h}_{lv}}V}
\end{equation}
where $\dot{\rho }$ is the volume-specific mass flux at the interface and $V$ is the cell volume. Eq.\ (\ref{eq:09a}) is calculated with the normal gradients computed at the liquid-vapor interface.

The energy equations for the fluid and solid regions are solved separately on different meshes connected by a common boundary at the solid-fluid interface. The energy conservation equation for the solid region is simply a heat conduction equation:
\begin{equation}\label{eq:09}
\frac{\partial ({{\rho }^{s}}c_{p}^{s}{{T}^{s}})}{\partial t}=\nabla \cdot ({{k}^{s}}\nabla {{T}^{s}})
\end{equation}
where ${{\rho }^{s}}$ and $c_{p}^{s}$ are the density and specific heat capacity of the solid material of the wall of the microchannels, respectively.

Due to the continuous nature of the temperature field and the heat flux at the solid-fluid interface, the temperature and the heat flux equations to handle the interfacial conditions between the solid and the fluid are, respectively:
\begin{equation}\label{eq:10b}
T_{\operatorname{int}}^{s}=T_{\operatorname{int}}^{f}
\end{equation}
\begin{equation}\label{eq:11b}
{{k}^{s}}\frac{\partial T_{\operatorname{int}}^{s}}{\partial n}={{k}^{f}}\frac{\partial T_{\operatorname{int}}^{f}}{\partial n}
\end{equation}

\subsection{Simulation setup}\label{sec:2.2}
\begin{figure}
  \centering
  \includegraphics[scale=0.6]{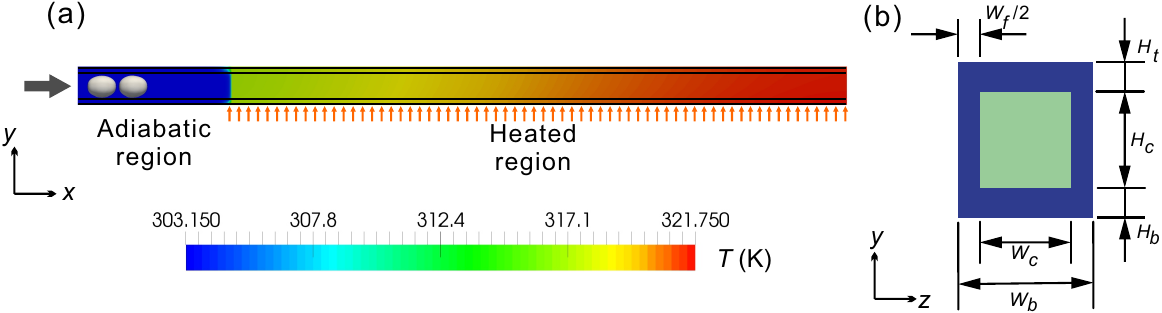}
  \caption{Simulation setup for multiple bubble flow boiling in microchannels. (a) Sectional view of the channel wall to show two bubbles and the internal temperature distribution in the channel. (b) Dimensions of the cross-section perpendicular to the $x$-direction. The blue region is the channel wall, and the green region is the working fluid.}\label{fig:01}
\end{figure}

	A square microchannel with a cross-section of $200 \times 200$ $\upmu$m$^2$ and a length of 6 mm is adopted in this work as shown in Figures \ref{fig:01}(a-b). The hydraulic diameter of the channel is $D_h = 4A/P = 200$ $\upmu$m, where $A$ and $P$ are the cross-sectional area and the perimeter of the microchannel. The microchannel is divided into two sections, an adiabatic section of length $L_a = 6D_h$ to allow the bubble to achieve a steady-state motion in the microchannel, and a heated section of length $L_h = 24D_h$, where a uniform wall heat flux is applied at the bottom base solid of the heated region. The top wall thickness, $H_t$, and vertical side wall thickness, $W_f$, are constant with $D_h/4$. Symmetry boundary conditions are imposed at the lateral surface of the domain to model the presence of adjacent channels, and only half ($W_f /2$) of the vertical side wall thickness is modeled. To reduce the computational cost, half of the computational domain is simulated with the symmetry condition at the middle cross-section in the $z$-direction.

	The flow dynamics and heat transfer of multiple bubbles during microchannel flow boiling with conjugated heat transfer are conducted in the microchannels by varying the number of bubbles, the initial bubble volume ratio, the liquid inlet Reynolds number, and the bottom wall thickness. The two-phase simulation begins with steady-state temperature and velocity fields obtained from single-phase simulations of liquid only at time $t = 0$. Each vapor bubble has a length of $1.1D_h$ and is initialized as a cylinder with spherical rounded caps and an initial axial diameter of $0.8D_h$, as shown in Figure \ref{fig:01} for a two-bubble case. The properties of the working fluid (R134a) and the microchannel wall (copper) are listed in Table 1. R134a is a suitable refrigerant for microchannel flow boiling due to its high dielectric strength, environmental friendliness, and low ozone depletion potential, making it ideal for diverse applications including high-power microelectronics. Copper offers exceptional thermal conductivity and machinability, allowing for efficient heat transfer and easy manufacturing. The velocity of the liquid is imposed at the inlet, while the temperature is set to $T_\text{sat}$. Unless otherwise specified, the liquid inlet velocity is $U = 0.31$ m/s with the Reynolds number $Re = \rho_l U_l D_h / \mu_l = 400$. A constant heat flux $q_w = 100$ kW/m$^2$ is imposed at the bottom base wall of the heated region.

\begin{table}[]
\centering
\caption{Properties of the working fluid (R134a) and the channel wall (copper) used in this study.}
\label{tab:01}
\begin{tabular}{lccc}
\hline
Properties                                         & R134a (Liquid)        & R134a (Vapor)          & Copper \\ \hline
Density, $\rho$ {[}kg/m$^3${]}                     & 1187.5                & 37.535                 & 8940   \\
Specific heat capacity, $c_p$ {[}J/(kg$\cdot$K){]} & 1446                  & 1065                   & 385    \\
Thermal conductivity, $k$ {[}mW/(m$\cdot$K){]}     & 80.27                 & 15.01                  & 398    \\
Dynamic viscosity, $\mu$ {[}mPa$\cdot$s{]}         & 0.1846                & 0.01238                & -      \\
Saturation temperature, $T_\text{sat}$ {[}K{]}     & \multicolumn{2}{c}{303.15}                     & -      \\
Surface tension, $\sigma$ {[}mN/m{]}               & \multicolumn{2}{c}{7.56}                       & -      \\
Latent heat, $h_{lv}$ {[}kJ/kg{]}                  & \multicolumn{2}{c}{173.1}                      & -      \\ \hline
\end{tabular}
\end{table}

The average Nusselt number is used to characterize the convection at the solid-fluid interface of the heated bottom wall ${{\overline{Nu}}_{\text{bottom}}}$, and it is calculated from the average convective heat transfer coefficient ${{\overline{h}}_{\text{bottom}}}$, 
\begin{equation}\label{eq:10}
{{\overline{Nu}}_{\text{bottom}}}=\frac{{{{\bar{h}}}_{\text{bottom}}}{{D}_{h}}}{{{k}_{l}}}
\end{equation}
\begin{equation}\label{eq:11}
{{\bar{h}}_{\text{bottom}}}=\frac{1}{A}\int_{0}^{A}{{{h}_{\text{bottom}}}dA}
\end{equation}
\begin{equation}\label{eq:12}
{{h}_{\text{bottom}}}=\frac{{{q}_{\operatorname{int}}}}{{{T}_{\mathrm{int}}}-{{T}_{\text{sat}}}}
\end{equation}
where ${{q}_{\operatorname{int}}}$ and $T_\text{int}$ are the local heat flux and local temperature at the solid-fluid interface of the heated bottom wall.
\subsection{Mesh independence study and validation}\label{sec:2.3}
\begin{figure}
  \centering
  \includegraphics[scale=0.3]{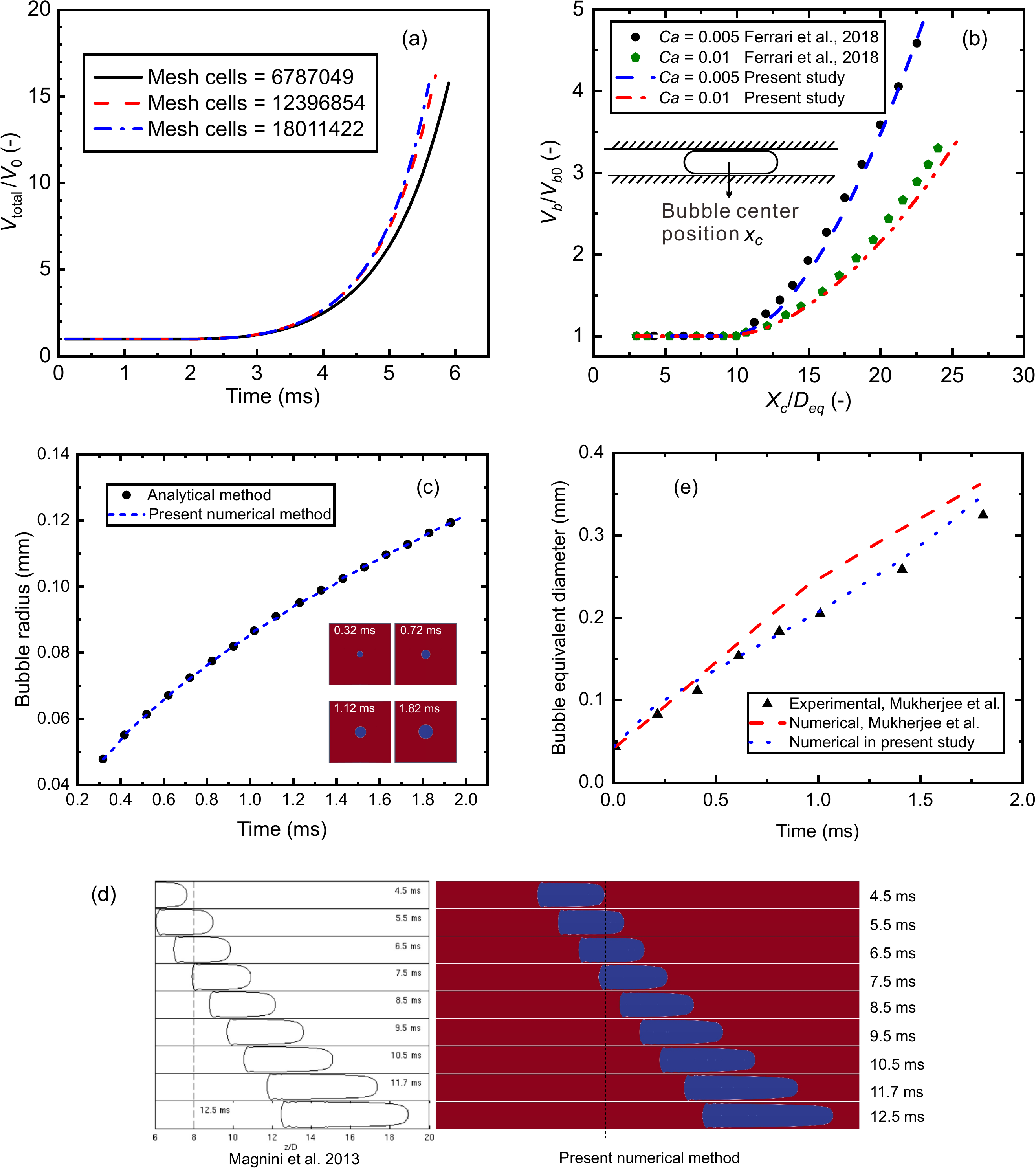}
  \caption{(a) Time variation of the total bubble volume $V_\text{total}/V_0$ for $H_b = 40$ $\upmu$m for a two-bubble case under three meshing schemes. (b) Dimensionless bubble volume, $V_b/V_{b0}$, as a function of the dimensionless bubble mass center position at different $Ca = \mu_l U_l /\sigma$ compared with the results obtained by Ferrari et al.\ \cite{Ferrari2018SlugFlowBoiling}.
  (c) Comparison of the variation of the bubble radius for spherical bubble growth in a superheated liquid with analytical result \cite{Scriven1959}.
  (d) Comparison of the snapshots of bubble growth during flow boiling for a case reported in Ref.\ \cite{Magnini2013ElongatedBubbleFlowBoiling} (Reprinted from Ref.\ \cite{Magnini2013ElongatedBubbleFlowBoiling} with permission from Elsevier).
  (e) Comparison of the bubble equivalent diameter with experimental results in Ref.\ \cite{mukherjee11}.
  }\label{fig:02}
\end{figure}

To balance the computational cost and accuracy, a mesh independence study is conducted. The grid cells are highly refined to resolve the bubble interface dynamics. Three different mesh schemes of about 6.8, 12.3, and 18.0 million cells are tested, with grid cell sizes of 2.373, 1.953, and 1.548 $\upmu$m, respectively, in the fluid domain. There is only a minute difference in the results between the mesh schemes of 12.3 and 18.0 million cells, as shown in Figure \ref{fig:02}(a), thus the meshing scheme with 12.3 million mesh cells is adopted.

The simulation is validated against the results of flowing bubble growth in a microchannel by Ferrari et al.\ \cite{Ferrari2018SlugFlowBoiling} for $Ca = 0.005$ and 0.01, which shows good agreement, as shown in Figure \ref{fig:02}(b).
The numerical model is also validated against the analytical benchmark problem of spherical bubble growth in superheated liquid. The evolution of the bubble radius shows good agreement with the analytical benchmark problem, as shown in Figure \ref{fig:02}(c). Further, the bubble growth during the microchannel flow boiling for a case reported in Ref.\ \cite{Magnini2013ElongatedBubbleFlowBoiling} is analyzed to validate the simulation and the results show good agreement with the bubble interface results of Magnini et al.\ \cite{Magnini2013ElongatedBubbleFlowBoiling} as shown in Figure \ref{fig:02}(d). Finally, we also validate our numerical model against the experimental results of flowing bubble growth in a microchannel by Mukherjee et al.\ \cite{mukherjee11}. As shown in Figure \ref{fig:02}(e), the computed bubble diameter agrees well with the experimental data. Therefore, these validations can prove the validity and accuracy of our numerical model.

\section{Results and discussion}\label{sec:3}
\subsection{Interaction between bubbles with different numbers of bubbles}\label{sec:3.1}

\begin{figure}
  \centering
  \includegraphics[scale=0.45]{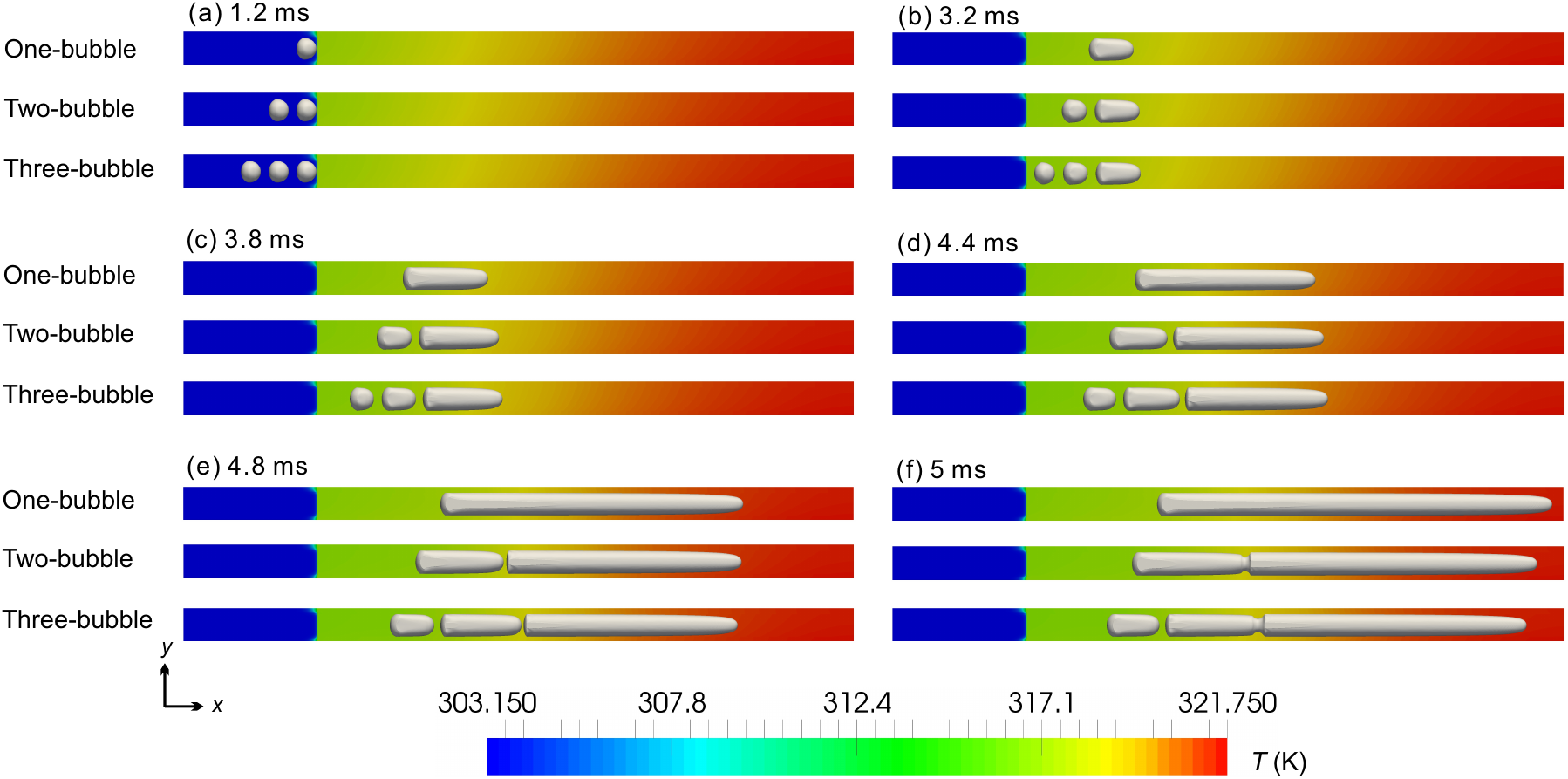}
  \caption{Snapshots of bubble growth and temperature fields at different instants for different numbers of bubbles. The temperature fields are the internal wall temperature obtained by cutting at the middle cross-section in the $z$-direction.}\label{fig:03}
\end{figure}

	The interaction between vapor bubbles during microchannel flow boiling is studied by varying the number of bubbles in the channels. Three cases are considered which comprise one, two, and three bubbles in the microchannels. The initial sizes of the bubbles are uniform for all the simulations with the same position of the leading bubbles. The bubbles’ growth and the solid domain internal temperature distribution for different numbers of bubbles are shown in Figure \ref{fig:03}. At the onset of the flow boiling process, the bubbles are transported in the flow direction by the saturated liquid that enters the microchannel.
Before the phase change process begins, the bubbles do not grow at the adiabatic section. Subsequently, when the bubbles enter the heated region, their sizes increase and their shapes change as they move downstream due to the phase change, as shown in Figure \ref{fig:03}. The bubbles’ respective sizes remain unchanged in the adiabatic region, as shown in Figure \ref{fig:03}(a), and start growing as their interfaces get in contact with the developed superheated thermal boundary layer at the solid-fluid interface of the microchannel heated section, as shown in Figure \ref{fig:03}(b).
For different numbers of bubbles in the microchannels, the sizes and positions of the leading bubbles are non-uniform during the flow boiling even though their initial sizes and positions are the same. 

\begin{figure}
  \centering
  \includegraphics[scale=0.45]{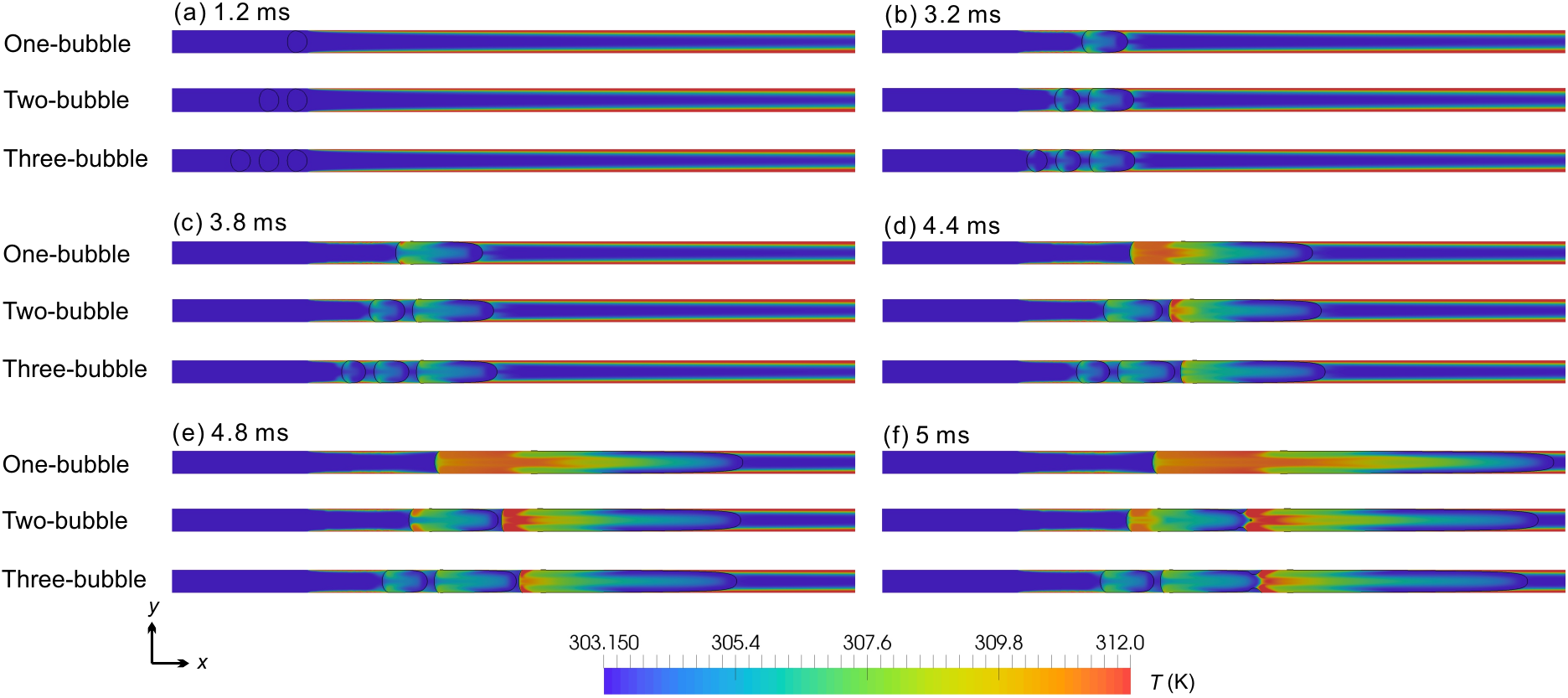}
  \caption{Temperature fields for microchannels with different numbers of bubbles taken at the middle cross-section in $z$-direction. Panels (a) show results when the bubbles are in the adiabatic region, while panels (b--f) show results when the bubbles are in the heated region.  
  }\label{fig:temperature}
\end{figure}

The leading bubble's growth rate decreases as the number of bubbles increases. This is because the vaporization at the rear bubbles absorbs a certain amount of heat, which reduces the local superheating of the fluid and reduces the expansion of the front bubble, as shown in Figure \ref{fig:temperature}. As the number of bubbles increases, the temperature at the rear of the leading bubble decreases. Therefore, the bubble size in the single-bubble case is much larger than that of the leading bubble in multiple-bubble cases. Further, as the adjacent trailing bubbles in multiple-bubble cases increase in size due to vaporization, the liquid slug separating the two consecutive bubbles reduces gradually, as shown in Figure \ref{fig:03}(b-f). 
Thus, as the trailing bubble gets closer to the rear end of the leading bubble, the liquid slug further reduces to a liquid film. The continuous expansion of the trailing bubble causes the thin liquid film to rupture and the bubbles to coalesce. The coalescence occurs as the trailing bubble growth rate increases and its nose has a higher velocity than the leading bubble’s rear velocity, thereby resulting in a long bubble that is elongated in the downstream as shown in Figure \ref{fig:03}(f).

\begin{figure}
  \centering
  \includegraphics[scale=0.45]{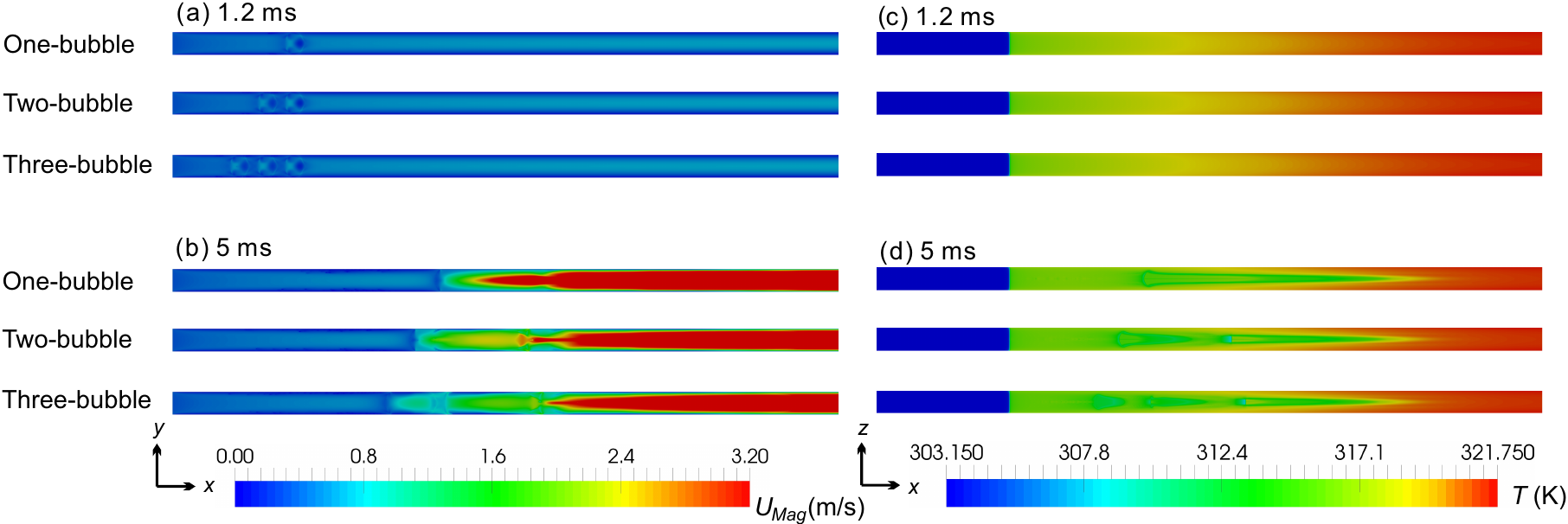}
  \caption{Flow boiling in microchannels with different numbers of bubbles: (a, b) velocity fields at the middle cross-section in the $z$-direction; (c, d) temperature fields at the solid-fluid interface of the bottom wall. Panels (a, c) show results when the bubbles are in the adiabatic region ($t = 1.2$ ms), while panels (b, d) show results when the bubbles are in the heated region ($t = 5$ ms).}\label{fig:04}
\end{figure}

\begin{figure}
  \centering
  \includegraphics[scale=0.3]{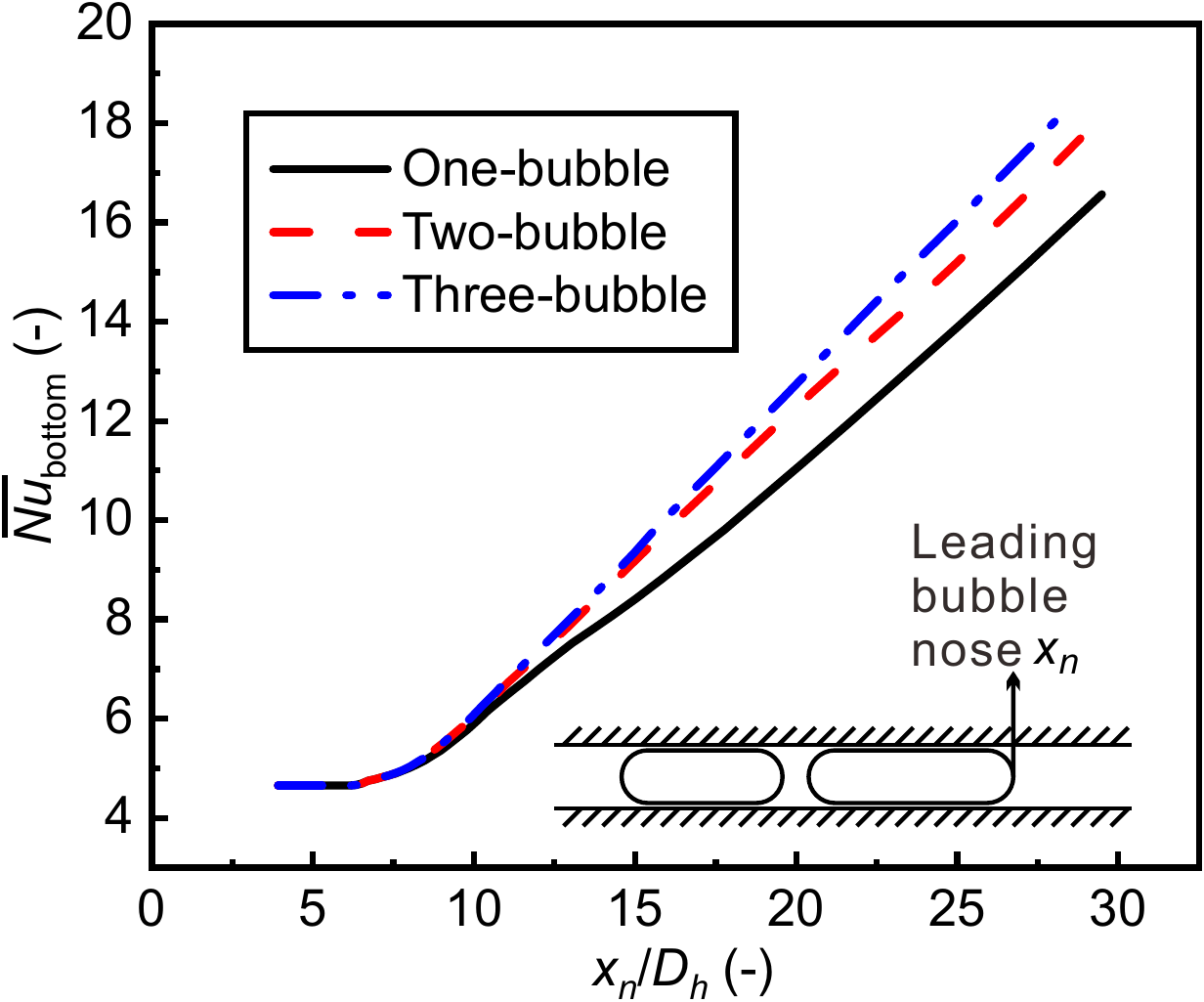}
  \caption{Nusselt number ${{\overline{Nu}}_{\text{bottom}}}$ plotted versus the dimensionless position of the leading bubble nose $x_n/D_h$ for different numbers of bubbles.}\label{fig:05}
\end{figure}

The flow velocity fields for different numbers of bubbles are shown in Figures \ref{fig:04}(a, b). The sequential flow with multiple bubbles produces stronger perturbation to the flow than that of a single vapor bubble.
The expansion of the trailing bubble pushes the liquid slug behind the leading bubble, causing the leading bubble to move faster compared to that of a single vapor bubble. Thus, the rear end of both the leading and trailing bubbles have non-uniform velocities for multiple-bubble microchannels (see the velocity distribution in the middle of the channel in Figure \ref{fig:04}). The velocity perturbation to the flow field by the multiple bubbles is higher compared to the perturbation caused by the single bubble, hence the convection from the wall to the fluid is enhanced in microchannels with multiple bubbles as shown by the Nusselt number in Figure \ref{fig:05}.

The velocity fields produced by the presence of the bubbles have a strong impact on the temperature field on the bottom wall, as shown in Figure \ref{fig:04}(c, d). When the bubbles are in the adiabatic region, the temperatures for different numbers of bubbles are almost the same, and the temperature on the bottom wall increases gradually along the flow direction. However, when the bubbles are in the heated section, the bubbles significantly perturb the flow fields, which then enhance the convection from the wall to the fluid. Therefore, the temperature on the bottom wall is remarkably stratified by the flow fields, producing a large temperature gradient in the spanwise direction. Because the velocity perturbation increases with the number of bubbles, the Nusselt number also increases correspondingly, as shown in Figure \ref{fig:05}.

\begin{figure}
  \centering
  \includegraphics[scale=0.3]{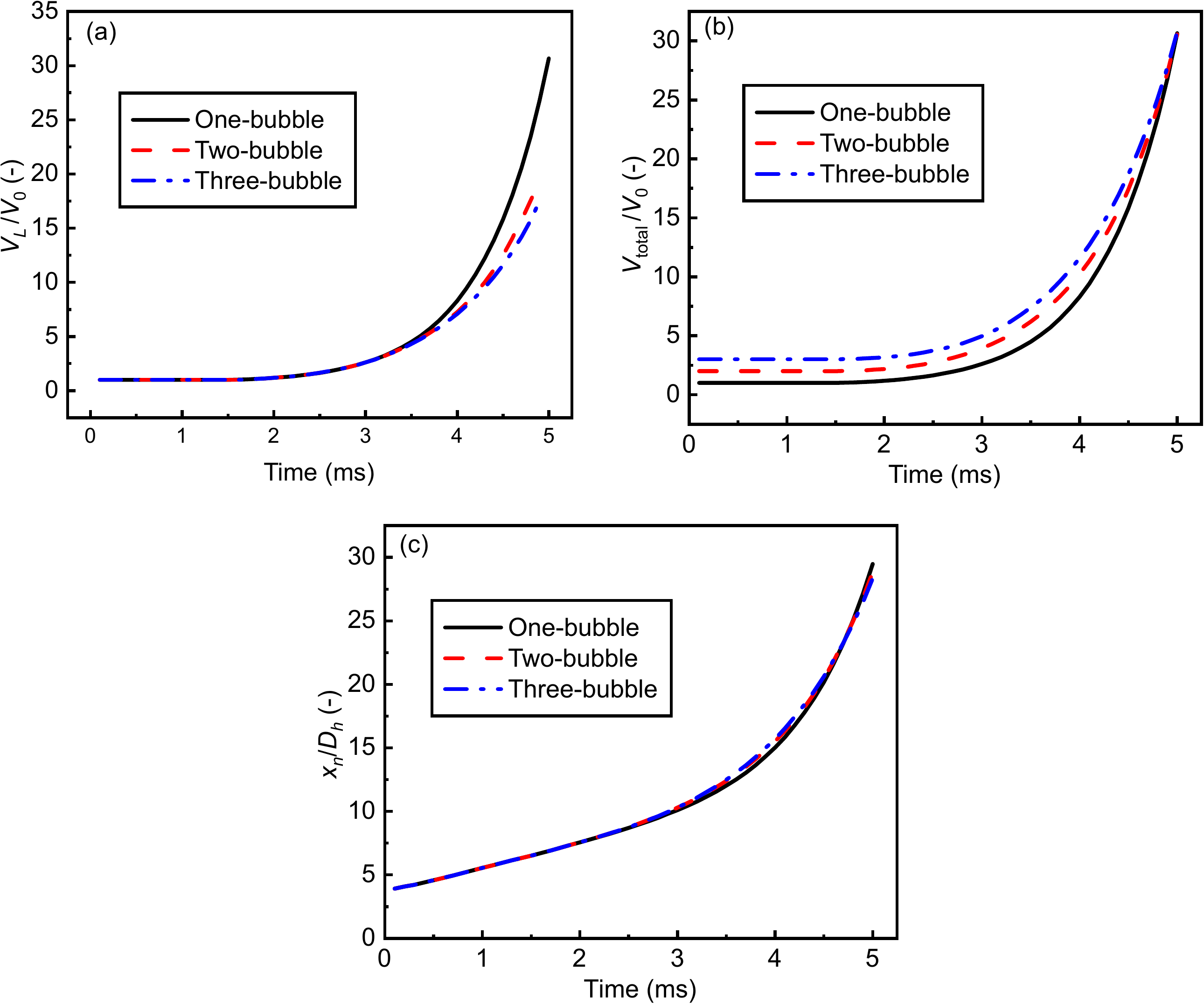}
  \caption{(a) Dimensionless leading bubble volume $V_L/V_0$, (b) dimensionless total bubble volume $V_\text{total}/V_0$, and (c) dimensionless position of the leading bubble nose $x_n/D_h$, plotted versus time for different numbers of bubbles.}\label{fig:06}
\end{figure}

The bubble growth for different numbers of bubbles in the microchannel are plotted as the dimensionless leading bubble volume, the dimensionless total bubble volume, and the leading bubble nose position in Figure \ref{fig:06}. The channel with the highest number of bubbles has the smallest leading bubble growth size as shown in Figure \ref{fig:06}(a). The increase in the number of bubbles has a significant impact on the leading bubbles’ growth size due to the heat absorption by the adjacent trailing bubbles, which reduces the degree of fluid superheating and also reduces the vaporization at the leading bubble. The total bubble volume is plotted in Figure \ref{fig:06}(b). Even though the initial total bubble volume increases with the number of bubbles, as the bubbles in the microchannels move downstream, the total volume gradually becomes more uniform. As a consequence, the leading bubble positions are not significantly affected by increasing the number of bubbles in the microchannel, as shown in Figure \ref{fig:06}(c). 

\begin{figure}
  \centering
  \includegraphics[scale=0.42]{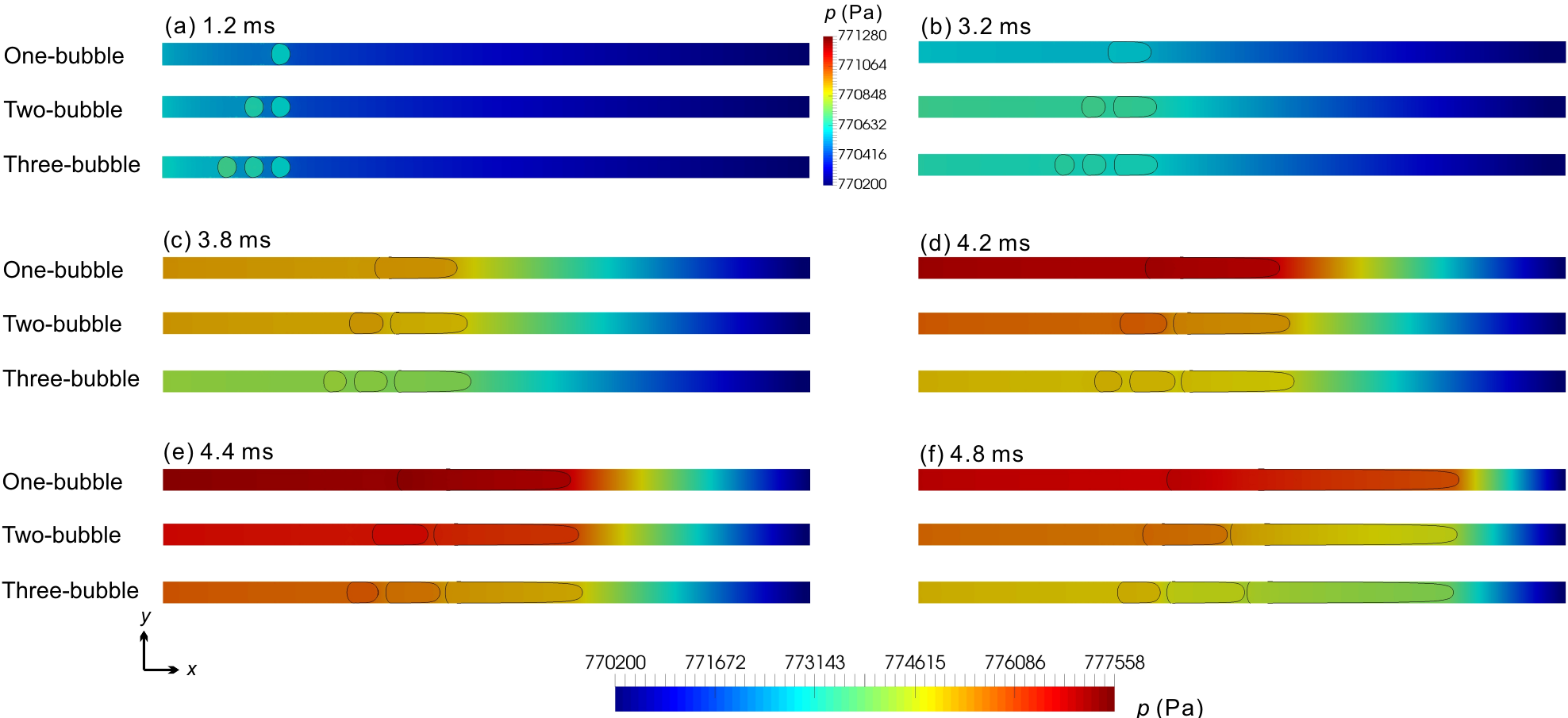}
  \caption{Pressure distribution contour at the middle cross-section in z-direction for microchannels with different numbers of bubbles.}\label{fig:pressure}
\end{figure}

The pressure distribution in the microchannels with different numbers of bubbles is presented in Figure \ref{fig:pressure}. The growth of the bubbles leads to variation in the pressure distribution. It is seen that a single-bubble microchannel tends to have a larger pressure drop compared to a multiple-bubble microchannel during the flow boiling process. This could be attributed to the presence of a large bubble causing the liquid in front to accelerate more than the leading bubble in multiple-bubble channels, leading to a pronounced pressure buildup.

\subsection{Effect of bubble volume ratio}\label{sec:3.2}
\begin{figure}
  \centering
  \includegraphics[scale=0.4]{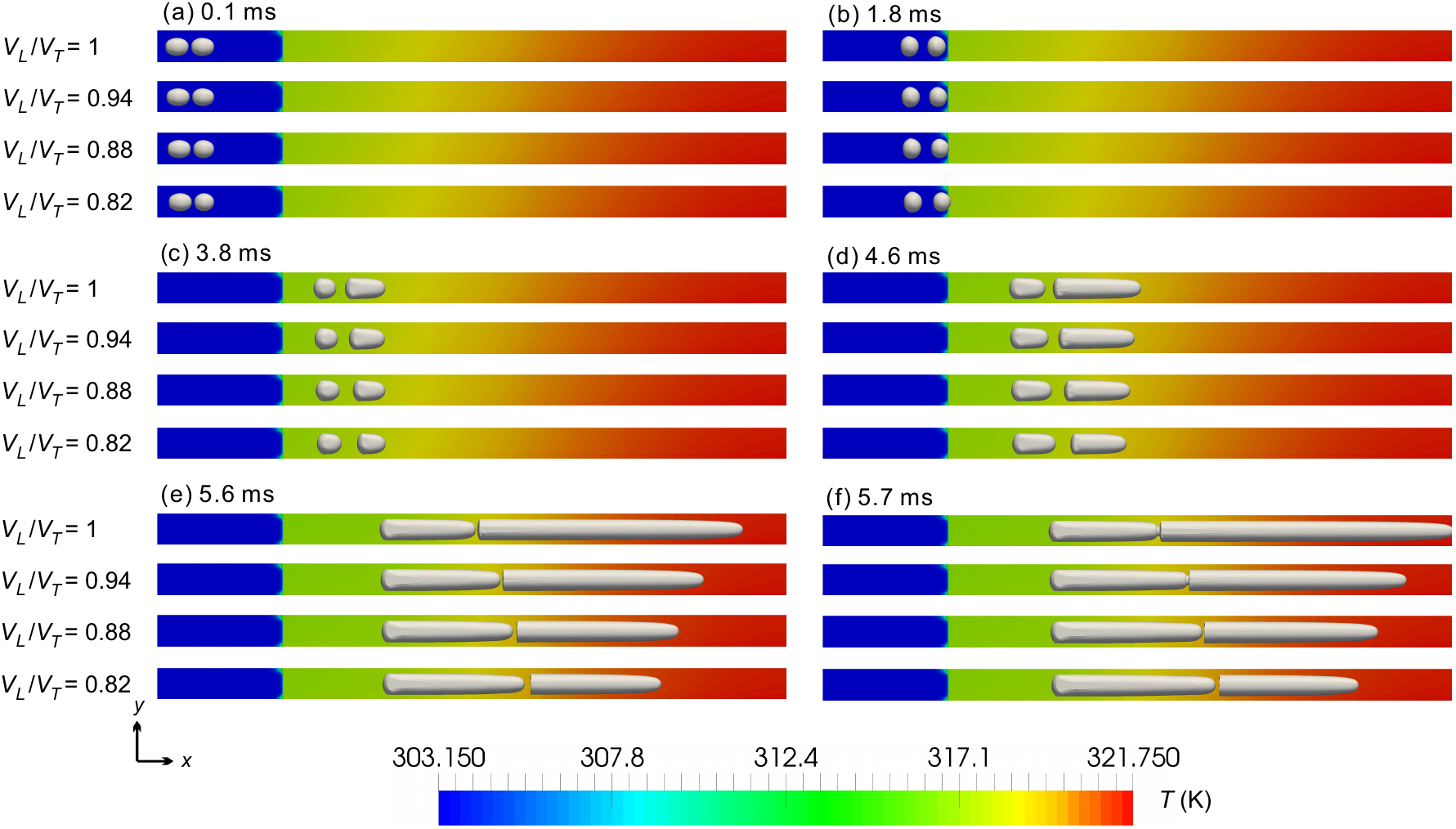}
  \caption{Snapshots of bubble growth and temperature fields at different instants for different bubble volume ratios. The temperature fields are the internal wall temperature obtained by cutting at the middle cross-section in the $z$-direction.}\label{fig:07}
\end{figure}

\begin{figure}
  \centering
  \includegraphics[scale=0.4]{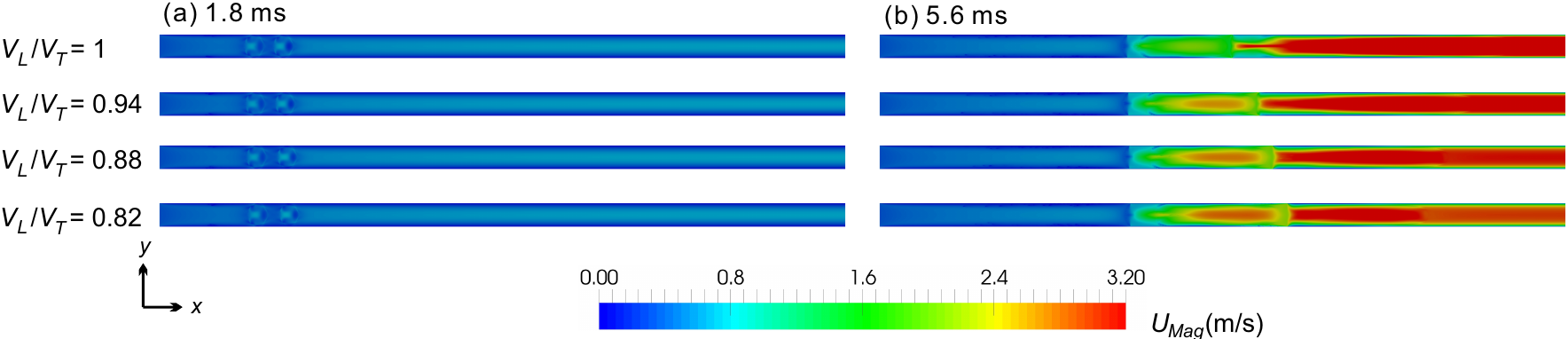}
  \caption{Velocity field at middle cross-section in the $z$-direction for different bubble volume ratios. Panel (a) shows results when the bubbles are in the adiabatic region ($t = 1.8$ ms), while panel (b) shows results when the bubbles are in the heated region ($t = 5.6$ ms).}\label{fig:08}
\end{figure}

\begin{figure}
  \centering
  \includegraphics[scale=0.3]{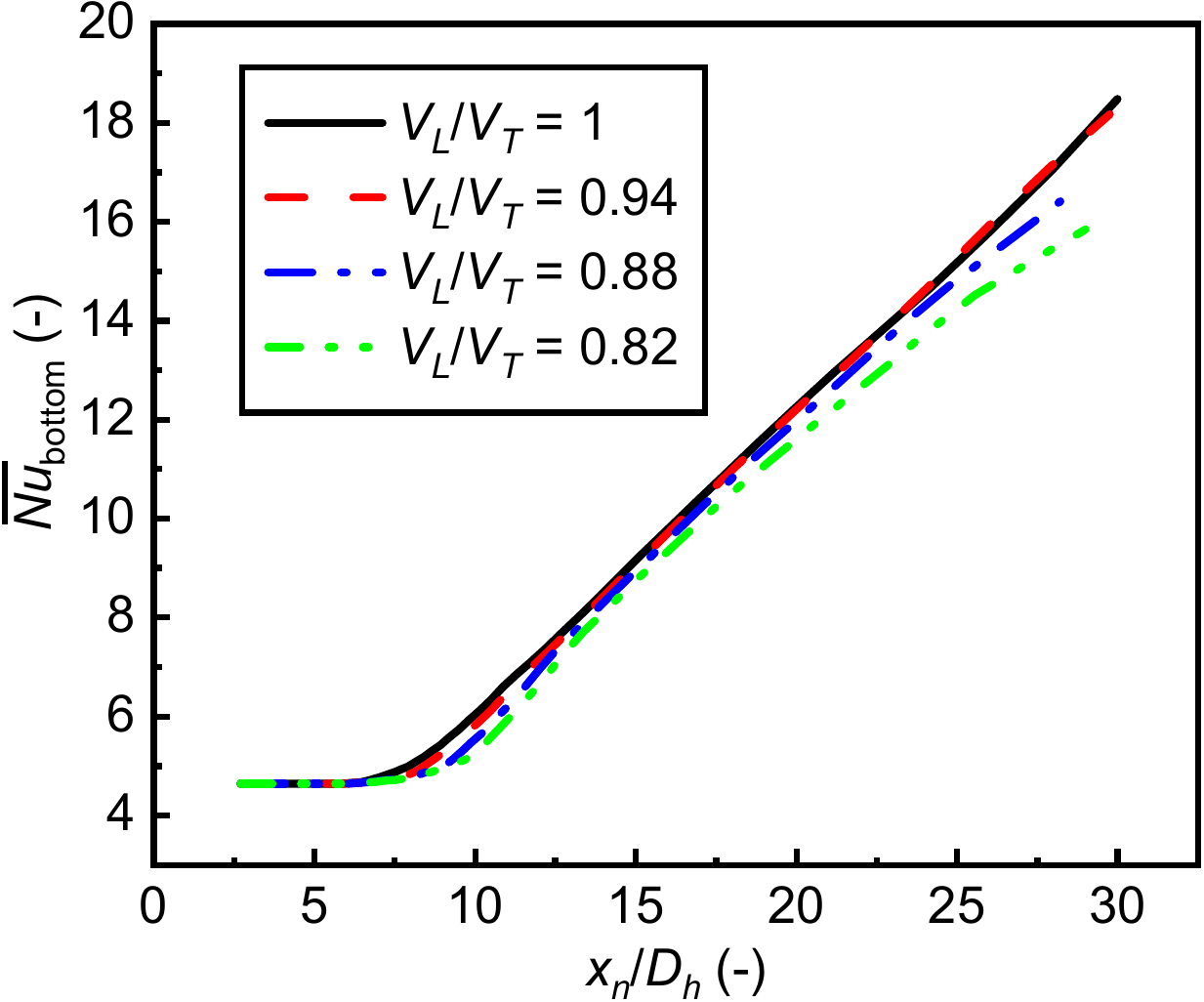}
  \caption{Nusselt number ${{\overline{Nu}}_{\text{bottom}}}$ plotted versus the dimensionless position of the leading bubble nose $x_n/D_h$ for different bubble volume ratios.}\label{fig:09}
\end{figure}

	To investigate the effects of the bubble volume ratio of two adjacent bubbles during microchannel flow boiling, the ratio of the leading bubble volume to the trailing bubble volume ($V_L/V_T$) is varied by changing the leading bubble volume $V_L$ while keeping the trailing bubble volume $V_T$ constant. Figure \ref{fig:07} shows the growth of the two vapor bubbles and the solid domain internal temperature distribution for different bubble volume ratios. The bubbles maintain their respective initial sizes when at the adiabatic region as shown in Figure \ref{fig:07} (a) and (b). In the heated region, as the initial size of the leading bubble decreases, its growth rate decreases. This is because the smaller initial bubble size reduces the contact with the superheated thermal boundary layer, leading to slower bubble growth compared to a larger leading bubble size. Meanwhile, as the leading bubble growth rate increases with the bubble volume ratio, the size of the trailing bubble decreases because the solid-fluid interface temperature is reduced after the passage of the leading bubble, thus reducing the vaporization rate for the trailing bubble. The result of the temperature field is also consistent with the flow fields where the perturbation to the flow by a smaller leading bubble decreases compared to a larger leading bubble as shown in Figure \ref{fig:08}. Consequently, the Nusselt number is reduced for smaller bubble volume ratios as shown in Figure \ref{fig:09}.

\begin{figure}
  \centering
  \includegraphics[scale=0.26]{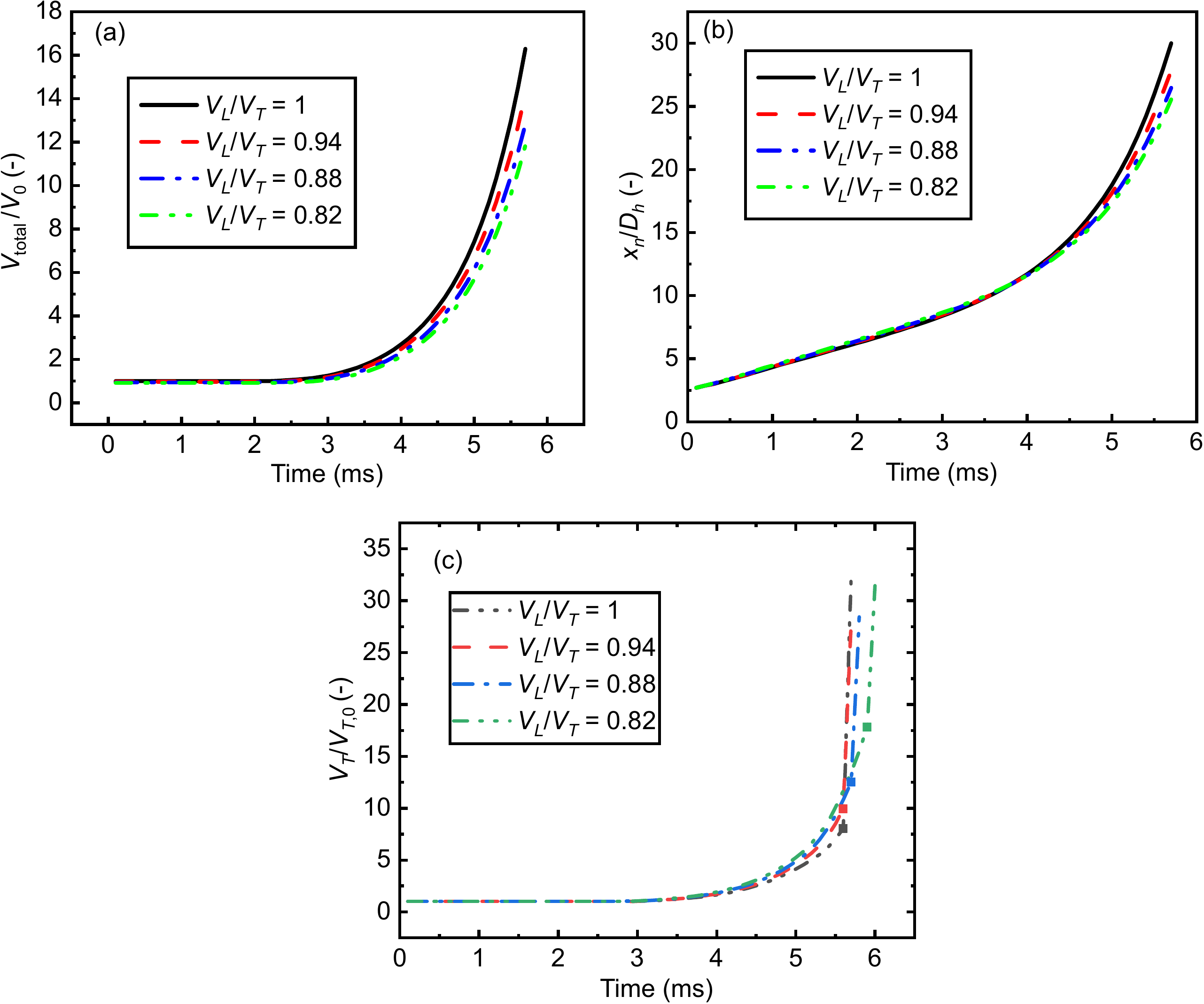}
  \caption{(a) Dimensionless total bubble volume $V_\text{total}/V_0$, (b) dimensionless position of the leading bubble nose $x_n/D_h$, and (c) dimensionless trailing bubble volume $V_T/V_{T,0}$, plotted versus time for different bubble volume ratios. The sudden increase of $V_T/V_{T,0}$ indicates the occurrence of the bubble coalescence and is marked with a square symbol. }\label{fig:10}
\end{figure}

To further evaluate the influence of the volume ratio of the leading bubble to the trailing bubble on the bubble dynamics, the dimensionless total bubble volume and the leading bubble nose position are plotted in Figures \ref{fig:10} (a) and (b), respectively. At the heated section, the channel with the largest bubble volume ratio has the largest total bubble growth and fastest speed, as shown in Figure \ref{fig:10}(a). This is not only because of the larger initial volume of the bubbles, but more importantly, because a larger initial bubble volume results in closer contact between the bubble and the thermal boundary layer, which further increases the expansion of the bubble due to vaporization. Regarding the bubble nose position, initially, at the adiabatic region, the bubble moves forward at a constant speed and the speed of the bubbles with smaller volume ratio is slightly higher than that of the larger volume ratio. This is because the fluid velocity in the center of the channel is higher than that near the wall, hence smaller bubbles, with less friction on the wall, move faster than larger bubbles \cite{odumosu23}. Afterward, in the heated region, the bubble nose accelerates rapidly because of the bubble expansion, and a larger bubble volume ratio leads to a faster acceleration because of the faster bubble expansion. The time variation of the bubble size of the trailing bubble is plotted in Figure \ref{fig:10}(c). We can note that the size of the trailing bubble suddenly increases at some instant in the late stage, indicating the coalescence of the two bubbles. Even though all the microchannels have the same initial trailing bubble size, the instants when the bubble coalescence occurs are different. A larger bubble volume ratio results in an earlier coalescence. The coalescence instants for the bubble volume ratios of 1 and 0.94 are very close at 5.6 ms, while for the bubble volume ratios of 0.88 and 0.82, the coalescence instants are 5.7 ms and 5.9 ms, respectively.

\subsection{Effect of inlet Reynolds number}\label{sec:3.3}
\begin{figure}
  \centering
  \includegraphics[scale=0.4]{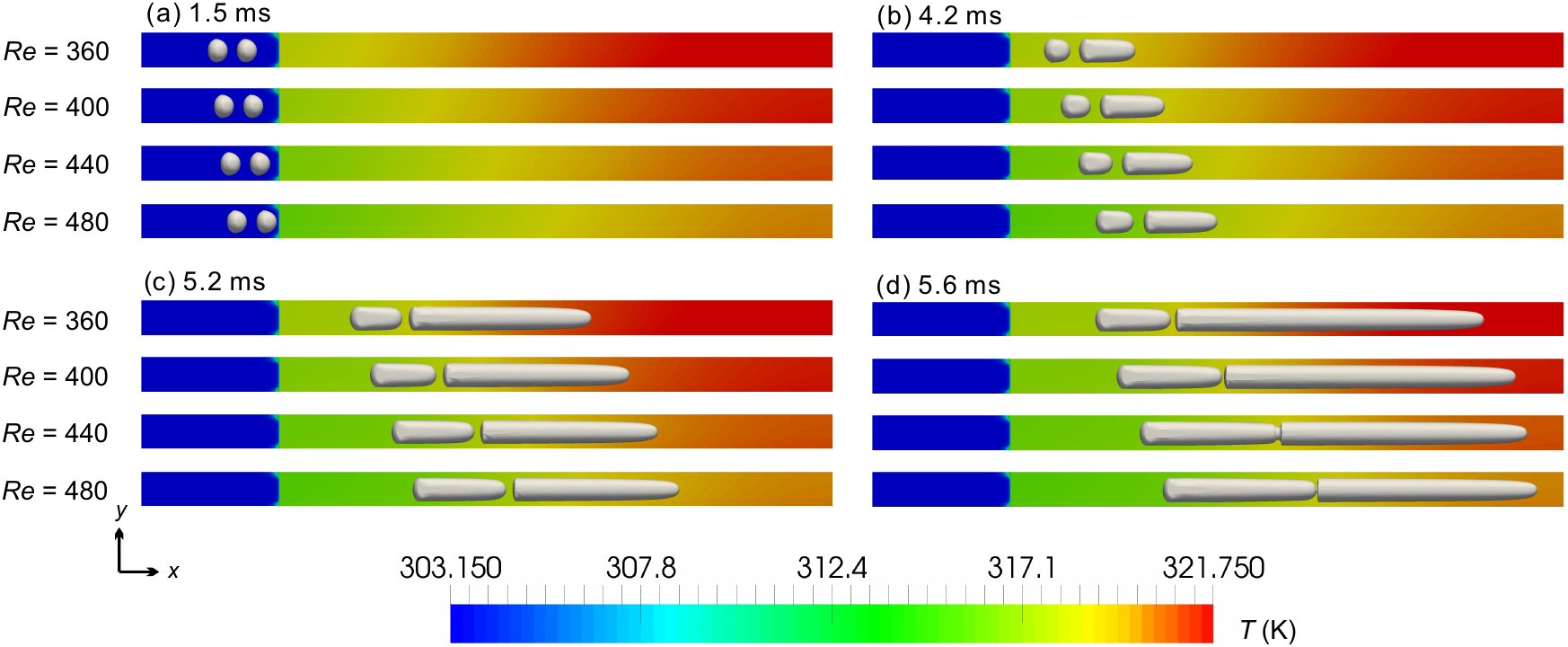}
  \caption{Snapshots of bubble growth and temperature fields at different instants for different $Re$ numbers for two-bubble cases. The temperature fields are the internal wall temperature obtained by cutting at the middle cross-section in the $z$-direction.}\label{fig:11}
\end{figure}

\begin{figure}
  \centering
  \includegraphics[scale=0.4]{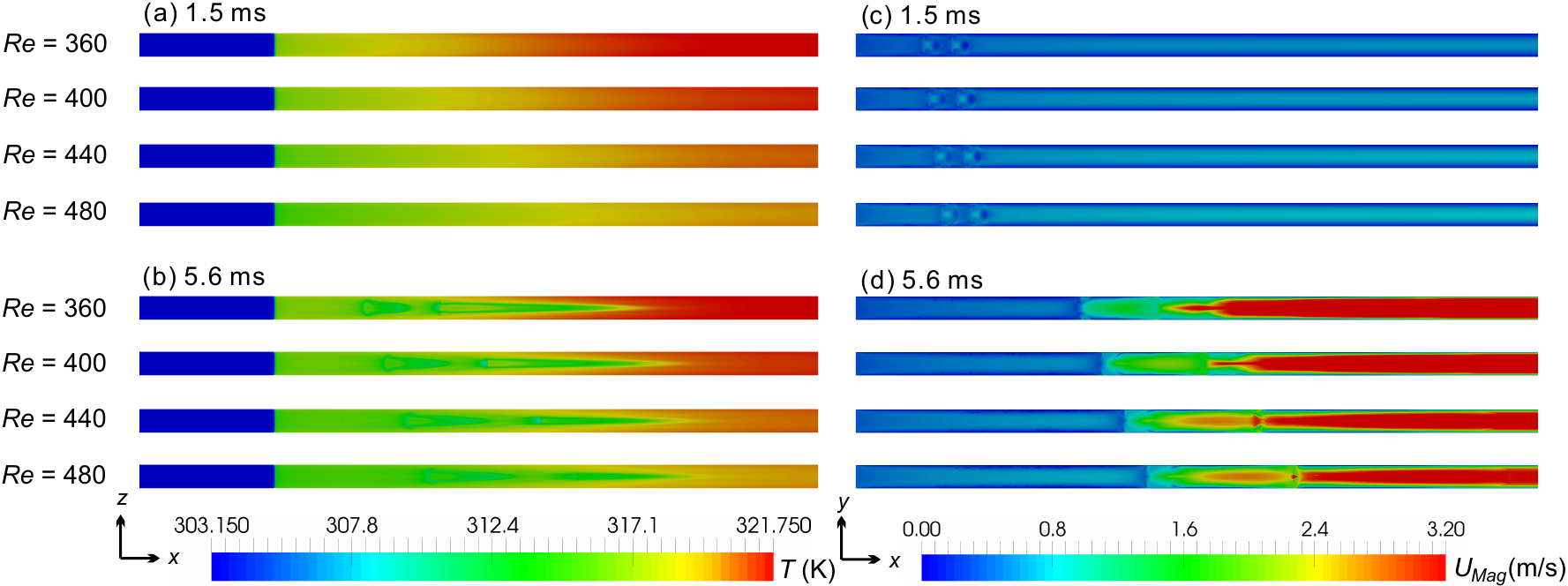}
  \caption{Flow boiling in microchannels with different $Re$ numbers for two-bubble cases: (a, b) temperature fields at the solid-fluid interface of the bottom wall; (c, d) velocity field at the middle cross-section in the $z$-direction. Panels (a, c) show results when the bubbles are in the adiabatic region ($t = 1.5$ ms), while panels (b, d) show results when the bubbles are in the heated region ($t = 5.6$ ms).}\label{fig:12}
\end{figure}

The influence of the inlet Reynolds number on multiple bubble growth is studied by changing the inlet liquid velocity. Figure \ref{fig:11} shows the solid domain internal wall temperature distribution and the multiple bubbles growth at different instants for different $Re$ numbers. In the adiabatic region, the respective bubbles’ sizes remain unchanged, and the temperature distribution of the microchannels at the heated region decreases linearly with increasing $Re$ number even when the same constant heat flux is applied at the bottom base as shown in Figure \ref{fig:11}(a). When the bubbles enter the heated region, the bubbles start growing and elongating downstream of the microchannel. The bubbles in the microchannel with the highest $Re$ number move fastest due to the high velocity of the inlet liquid. At the upstream of the heated region, when the bubbles start growing in the microchannel, both the leading and the trailing bubbles of the microchannels increase slightly in size with increasing the $Re$ number (see Figure \ref{fig:11}b). This is because the bubbles in the microchannels with higher $Re$ numbers move faster and firstly get in contact with the superheated fluid in the heated region, which also increases the bubble growth rate. However, after a while, a new trend is established whereby, the leading bubble size increases as $Re$ decreases (see Figure \ref{fig:11}d). The leading bubble of the microchannel with the lowest $Re$ number has the largest leading bubble size. This can be attributed to the higher temperatures exhibited at the downstream of the microchannel with low $Re$ numbers as shown in Figures \ref{fig:12}(a, b), which thickens the thermal boundary layer, subsequently leading to the increase in the bubble size and enhancing the heat transfer. 

The flow fields in microchannels with different $Re$ numbers are shown in Figure \ref{fig:12} (c, d). The perturbation to the flow field produced by the multiple bubbles enhances the thermal convection from the wall to the fluid. The perturbations produced by the leading bubbles of different $Re$ numbers are larger than that of the trailing bubble, but the effects are non-uniform at different $Re$ numbers. As the $Re$ number increases, the leading bubble perturbation decreases while the adjacent trailing bubble perturbation increases. As the leading bubble size grows larger than the trailing bubble, the difference in the velocity between the leading and trailing bubbles further increases. Because the microchannels with lower $Re$ numbers produce larger leading bubbles, hence the large leading bubbles significantly perturb the flow in the microchannel. 

\begin{figure}
  \centering
  \includegraphics[scale=0.3]{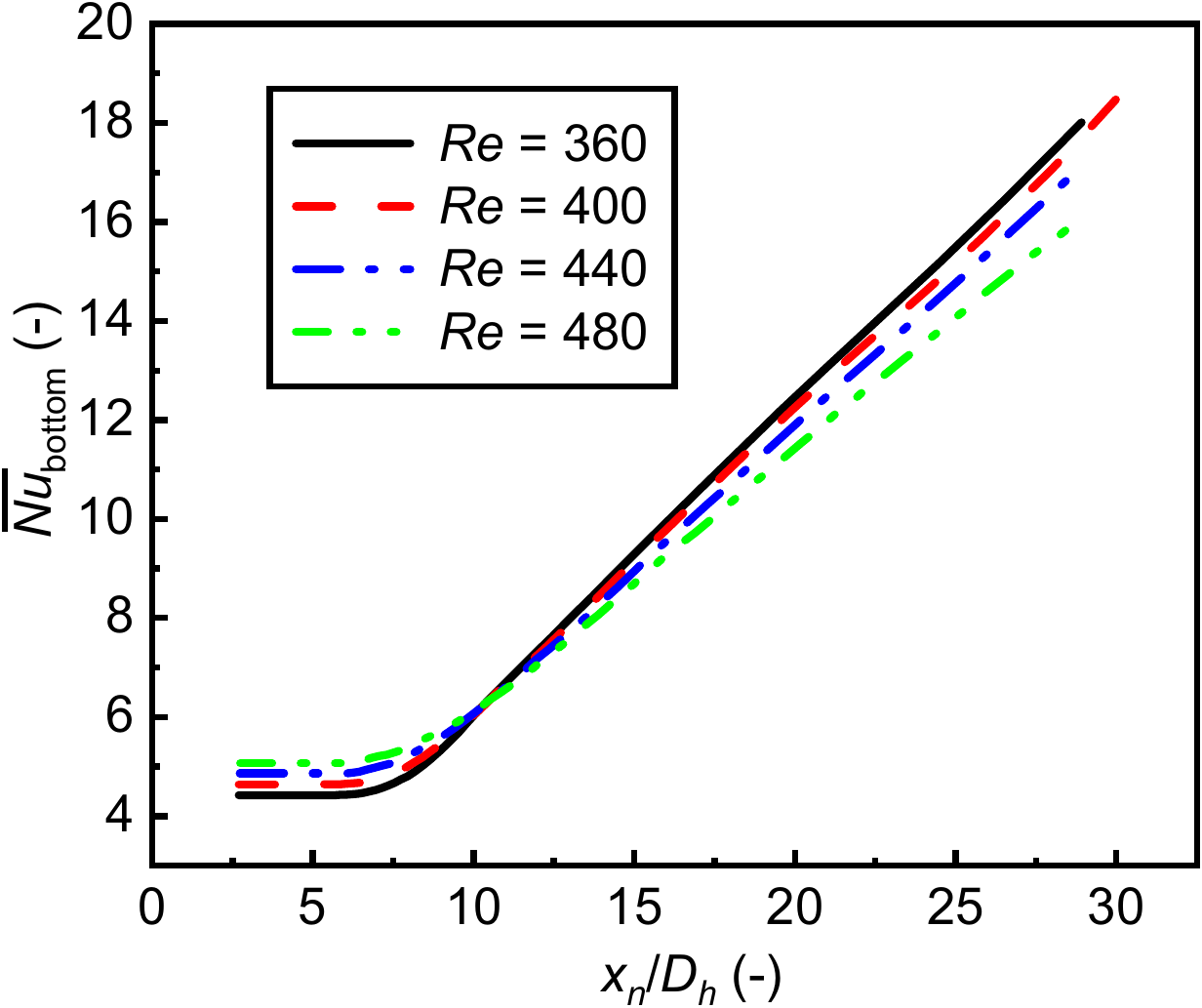}
  \caption{Nusselt number ${{\overline{Nu}}_{\text{bottom}}}$ plotted versus the dimensionless position of the leading bubble nose $x_n/D_h$ for different $Re$ numbers.}\label{fig:13}
\end{figure}

Because of the different trends of the bubble growth behaviors in the upstream and downstream as $Re$ increases, the Nusselt number also shows different trends in the upstream and downstream, as shown in Figure \ref{fig:13}. When the bubbles are in the upstream of the heated region, the Nusselt number increases with $Re$. This trend is reversed when the bubbles move to the downstream of the heated region. This is a consequence of the different bubble moving speeds in different regions. In the upstream of the heated region, the moving speed of the bubble is mainly determined by the inlet Reynolds number of the flow. Hence, a higher Reynolds number results in a higher bubble moving speed, a stronger effect of convection, and a higher Nusselt number. In contrast, when the bubble is in the downstream of the heated region, the moving speed of the bubble is mainly determined by the bubble expansion because of the rapid expansion of the two bubbles under vaporization. Hence, a higher Reynolds number results in a lower superheat degree, a slower expansion of the bubble size, and a lower Nusselt number. 

\begin{figure}
  \centering
  \includegraphics[scale=0.3]{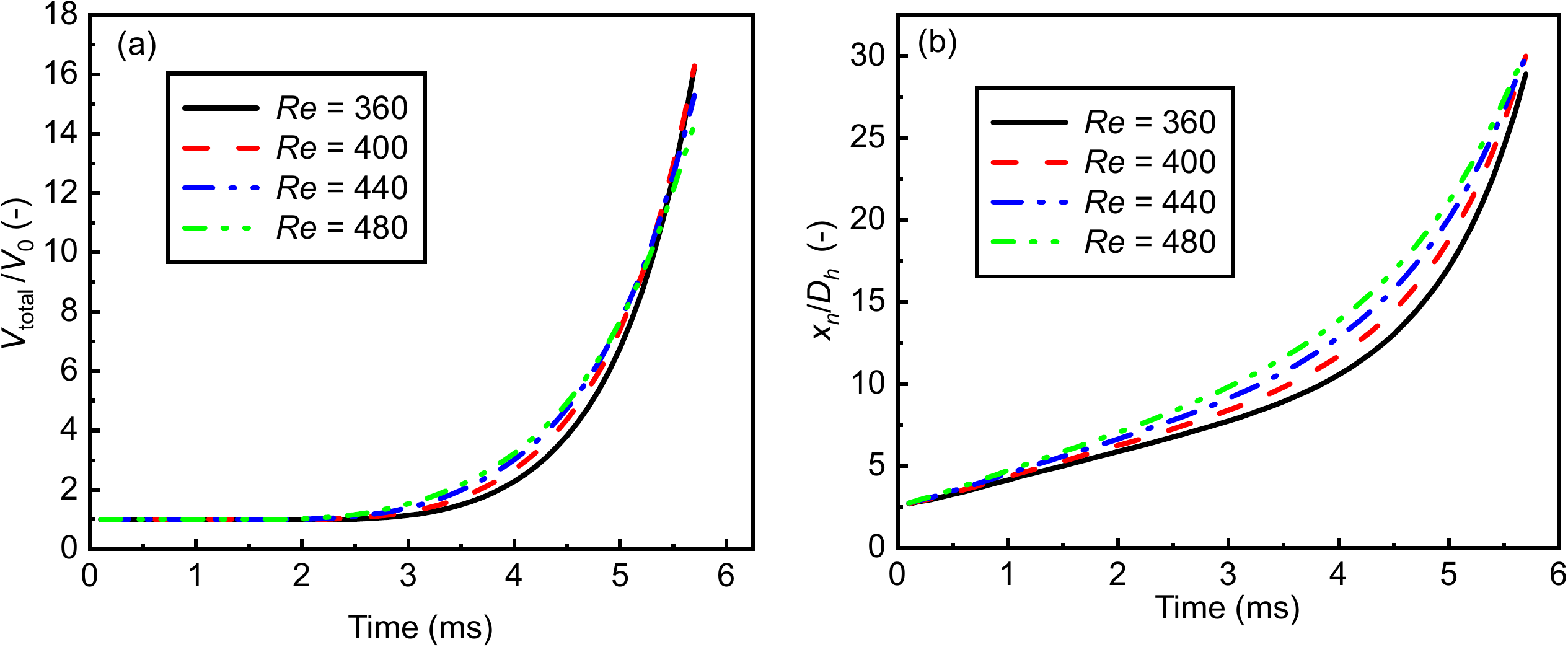}
  \caption{(a) Dimensionless total bubble volume $V_\text{total}/V_0$, (b) dimensionless position of the leading bubble nose $x_n/D_h$, plotted versus time for different $Re$ numbers.}\label{fig:14}
\end{figure}

To further evaluate the influence of the inlet $Re$ numbers on the bubble dynamics, the dimensionless total bubble volume and the leading bubble nose position for different $Re$ numbers are plotted in Figures \ref{fig:14} (a) and (b), respectively. In the upstream of the heated section, as the $Re$ number increases, the total bubble volume increases. However, when the bubbles move to the downstream of the microchannel, the bubble volumes at different $Re$ numbers become more uniform, as shown in Figure \ref{fig:14}(a). The difference in the bubble volume in the upstream is mainly because the bubble is pushed by the fluid at different inlet speeds, i.e., the bubbles in higher $Re$ numbers move faster. However, in the downstream of the microchannel, the bubbles at lower $Re$ numbers expand more quickly because of the higher superheat degree of the fluids. Therefore, the effect of the slower inlet velocity and the faster bubble expansion can cancel the overall effect of bubble expansion. Hence, the bubble volume at small $Re$ numbers can catch up with the bubble size at high $Re$ numbers or even become larger. This trend is also reflected in the curves of the bubble nose position in Figure \ref{fig:14}(b). Initially, the bubbles at low $Re$ numbers move slower than those at high $Re$ numbers, but they can quickly catch up in the later stage because of the bubble expansion. 

\subsection{Effect of bottom wall thickness}\label{sec:3.4}
\begin{figure}
  \centering
  \includegraphics[scale=0.4]{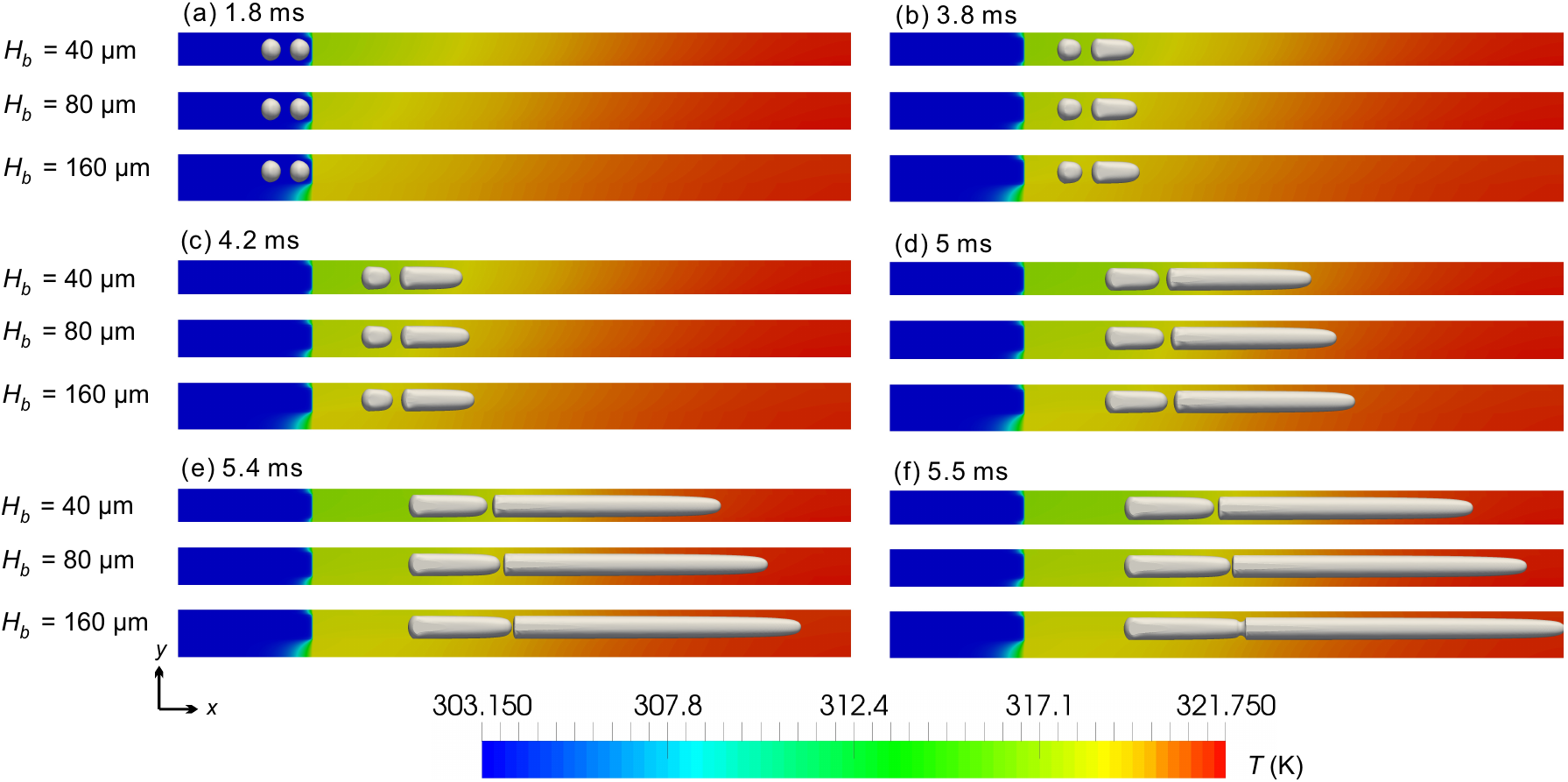}
  \caption{Snapshots of bubble growth and temperature fields at different instants for different bottom wall thicknesses. The temperature fields are the internal wall temperature obtained by cutting at the middle cross-section in the $z$-direction.}\label{fig:15}
\end{figure}

\begin{figure}
  \centering
  \includegraphics[scale=0.26]{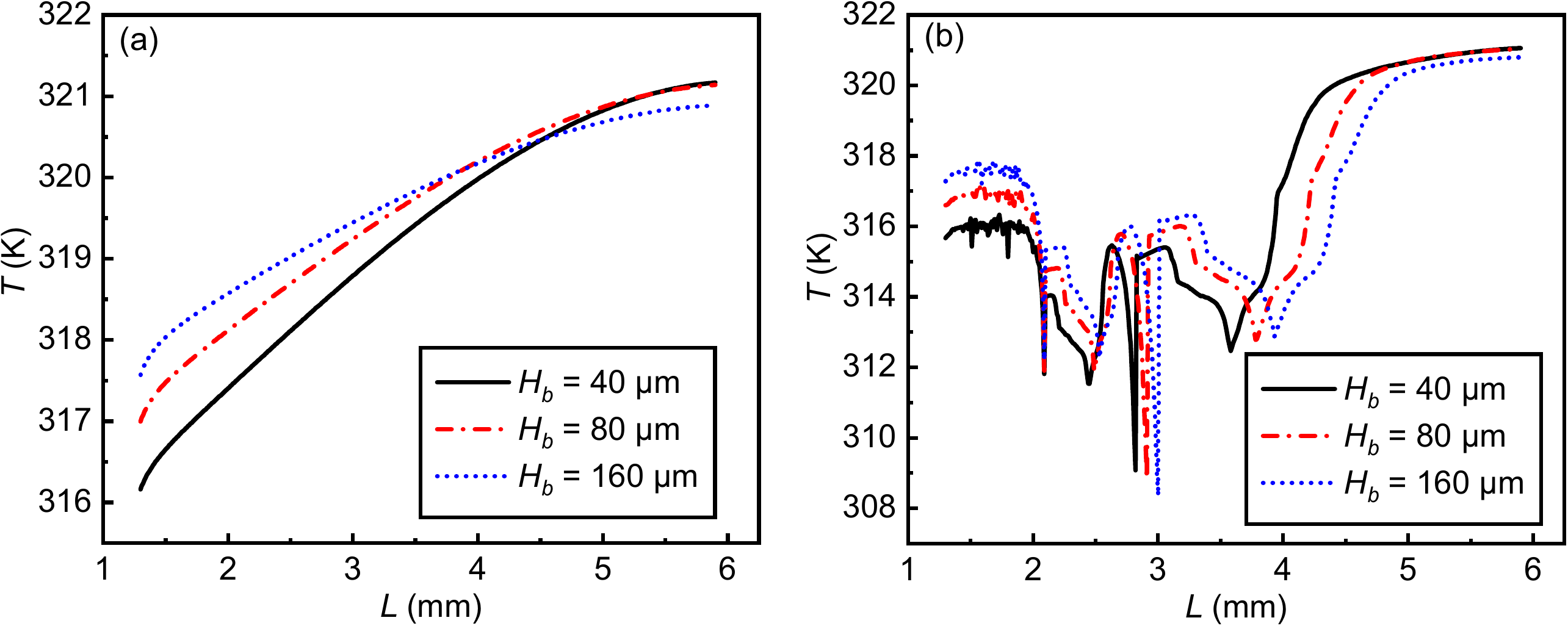}
  \caption{Temperature distribution for different bottom wall thicknesses along the centre line of the solid-fluid interface of the heated bottom wall ($z = 0$). (a) Temperature distribution when the bubbles are in the adiabatic region ($t = 1.8$ ms), (b) Temperature distribution when the bubbles are in the heated region ($t = 5.4$ ms).}\label{fig:16}
\end{figure}

\begin{figure}
  \centering
  \includegraphics[scale=0.4]{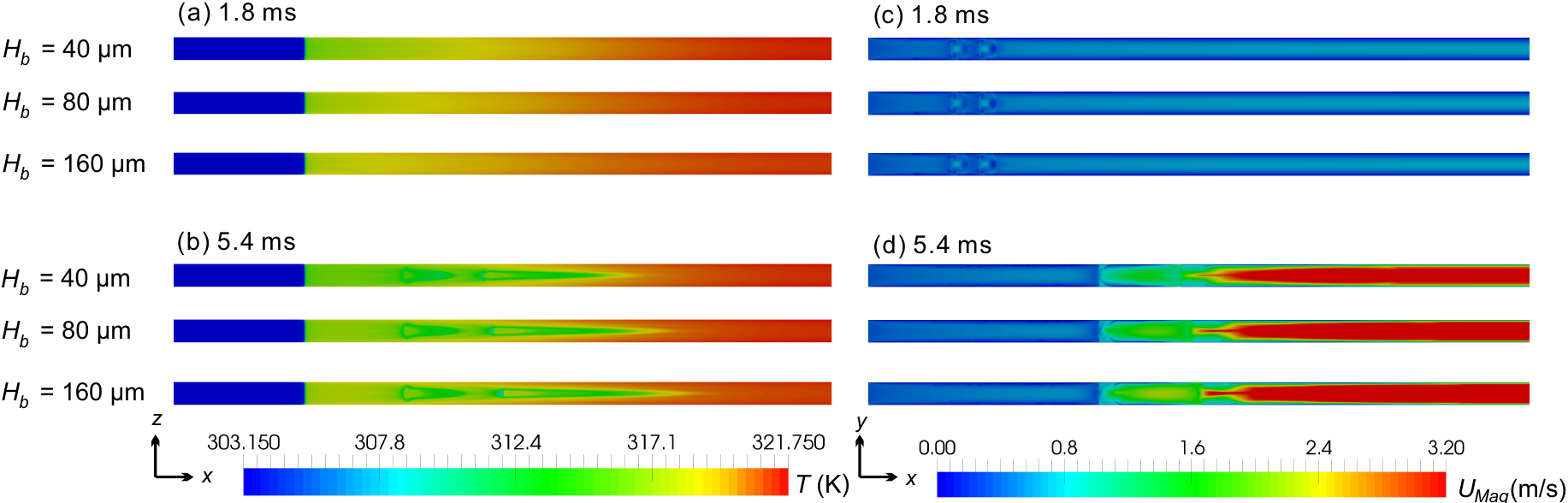}
  \caption{Flow boiling in microchannels with different bottom wall thicknesses for two-bubble case: (a, b) temperature fields at the solid-fluid interface of the bottom wall; (c, d) velocity field at the middle cross-section in the $z$-direction. Panels (a, c) show results when the bubbles are in the adiabatic region ($t = 1.8$ ms), while panels (b, d) show results when the bubbles are in the heated region ($t = 5.4$ ms).}\label{fig:17}
\end{figure}

In practical applications, varying the bottom wall thickness determines the microchannel compactness, the ease of manufacturing process, and material limitations, which is crucial for the design and optimization of the microchannel heat sinks.
The influence of the bottom wall thickness on the dynamics of multiple bubbles during microchannel flow boiling is studied by varying the bottom wall thicknesses, $H_b$. Even though the initial bubble sizes are the same at the adiabatic region, their respective sizes are non-uniform during the flow boiling for different bottom wall thicknesses, as shown in Figure \ref{fig:15}. The leading bubble becomes larger than the trailing bubbles because of its contact with higher superheat temperature at the solid-fluid interface of the microchannel. After the passage of the leading bubble through the channel, the superheat temperature behind it is reduced, and the trailing bubble size is smaller than the leading bubble due to its contact with lower local temperature. As the bottom wall thickness increases, the leading bubble size increases faster, and becomes larger compared to the leading bubble of the thinner bottom wall. This is because, with a thicker bottom wall, the heated wall temperature is more uniform along the flow direction because of the heat conduction in the solid bottom wall, hence producing a higher temperature in the upstream compared to that with a thinner bottom wall, as shown in Figures \ref{fig:16}(a) and \ref{fig:17}(a). Therefore, the vaporization at a thicker bottom wall is faster and the bubble size becomes larger. Similarly, the trailing bubble also increases with the bottom wall thickness due to the same mechanism. As a consequence of the size increase of both bubbles due to increased vaporization, the liquid slug separating the two bubbles reduces gradually. Thus, the liquid slug reduces to a liquid film, which later ruptures, causing the coalescence of the two bubbles, as shown in Figure \ref{fig:15}(f). Due to the different bottom thicknesses, the instants of the bubble coalescence are not the same. For microchannels with thicker bottom walls, the bubbles coalesce earlier than microchannels with thinner bottom walls.

The flow velocity fields for different bottom wall thicknesses are shown in Figure \ref{fig:17}(c, d). The perturbation to the flow by the two consecutive bubbles increases as the bottom wall thickness increases. The expansion of the trailing bubble causes the acceleration of the liquid slug, which further pushes the leading bubble’s rear end. Thus, the rear ends of the leading bubbles for different wall thicknesses are non-uniform. The velocity perturbation to the flow field in microchannels with thicker bottom walls is higher than that with thinner bottom thickness, hence the convection from the wall to the fluid is enhanced in microchannels with thicker bottom walls.

\begin{figure}
  \centering
  \includegraphics[scale=0.26]{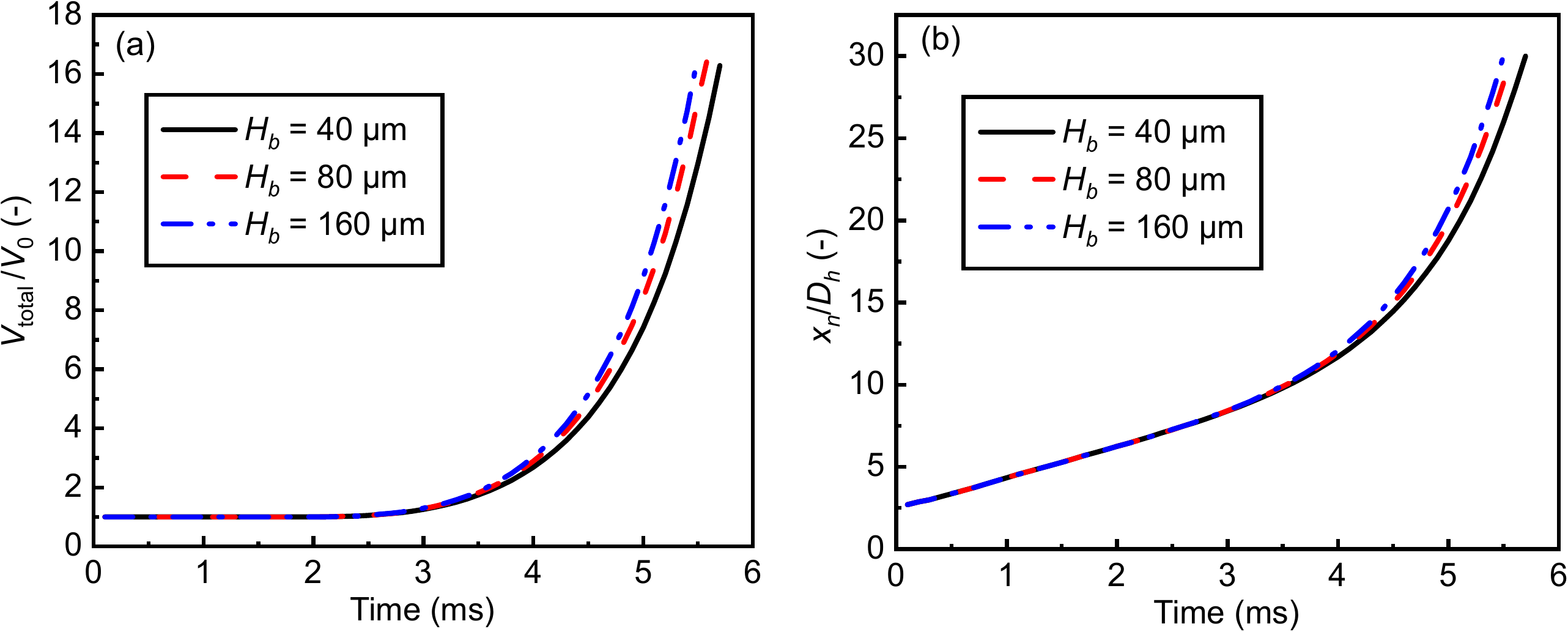}
  \caption{(a) Dimensionless total bubble volume $V_\text{total}/V_0$, (b) dimensionless position of the leading bubble nose $x_n/D_h$, plotted versus time for different bottom wall thicknesses.}\label{fig:18}
\end{figure}

The influence of the bottom wall thickness on the bubble dynamics is quantified by plotting the dimensionless total bubble volume and the leading bubble nose position in Figures \ref{fig:18}(a, b). The channels with thicker bottom walls have larger bubble sizes and faster moving speeds. This is because the variation of the bottom wall thickness influences the bubbles’ growth and acceleration. The bubble acceleration is caused by the expansion and interaction of the two bubbles while moving forward due to the phase change. 

\begin{figure}
  \centering
  \includegraphics[scale=0.3]{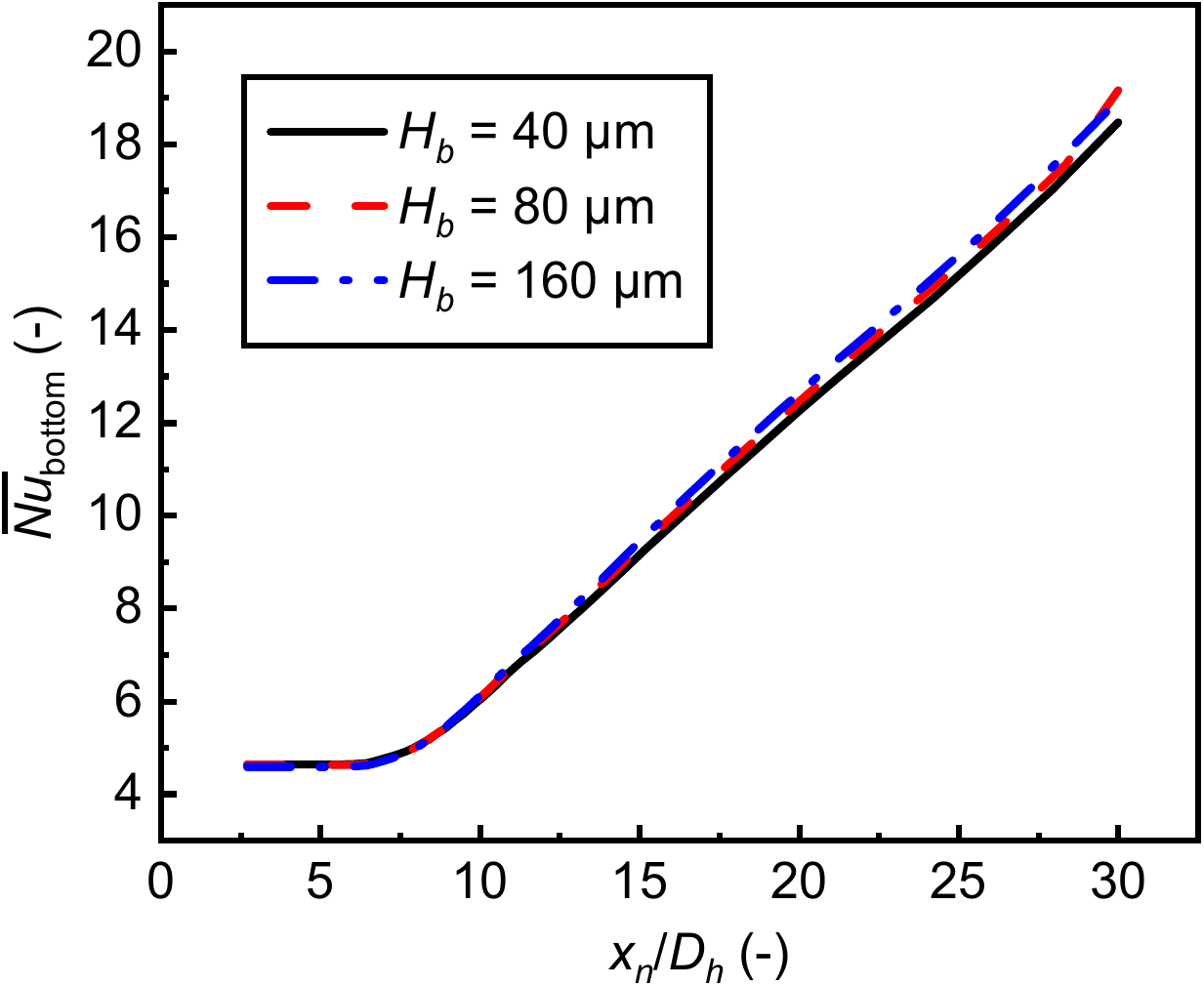}
  \caption{Nusselt number ${{\overline{Nu}}_{\text{bottom}}}$ versus the dimensionless position of the leading bubble nose $x_n/D_h$ for different bottom wall thicknesses.}\label{fig:19}
\end{figure}

The bottom wall thickness also affects the Nusselt number for the multiple-bubble flow in microchannels. The average Nusselt number at the heated bottom solid-fluid interface of the microchannel ${{\overline{Nu}}_{\text{bottom}}}$ is plotted versus the dimensionless leading bubble nose position $x_n/D_h$ in Figure \ref{fig:19}. The Nusselt number is small and constant initially when the bubble is at the adiabatic region. The Nusselt number quickly increases as vaporization begins at the heated section. The microchannels with thicker bottom walls have better heat transfer performance compared to those with thinner bottom walls. The difference in the magnitude of the Nusselt number is small because the microchannel material adopted (copper) has both higher thermal conductivity and diffusivity, which make it to transfer heat effectively. 

\section{Conclusions}\label{sec:4}
	In this study, the interaction between vapor bubbles in flow boiling heat transfer in microchannels is investigated numerically. The significant impacts that the number of bubbles, bubble volume ratio, inlet Reynolds number, and wall thickness play on the flow dynamics and heat transfer are emphasized. For different numbers of bubbles in the microchannels, the sizes and positions of the leading bubbles are non-uniform during the flow boiling even when their initial sizes and positions are the same. The size of the bubble in a single-bubble microchannel is larger compared to the leading bubble of multiple-bubble cases because the vaporization at the rear bubbles absorbs heat and reduces both the local superheating of the fluid and the expansion of the leading bubble. The increase in the number of bubbles within the channel also increases the Nusselt number of heat transfer because of the flow perturbation by the bubbles. As the initial volume ratio between the leading bubble and the rear bubble decreases, the leading bubble size in the downstream becomes smaller because the smaller initial size reduces the contact with the superheated thermal boundary layer; meanwhile, the size of the trailing bubble increases because, with less heat absorption by the leading bubble, the fluid surrounding the trailing bubble has a higher temperature. The initial volume ratio between the leading bubble and the rear bubble also affects the instant of bubble coalescence, and a larger initial volume ratio results in an earlier coalescence of the two bubbles. With increasing $Re$, both the leading and the trailing bubbles increase slightly in size in the upstream of the heated region, because the bubbles at higher $Re$ move faster and firstly get in contact with the superheated fluid; however, in the downstream, the leading bubble size increases with decreasing $Re$ because of the higher temperature at the solid-fluid interface of the microchannel at low $Re$. At the upstream, a higher Reynolds number results in a higher bubble moving speed, a stronger effect of convection, and a higher Nusselt number, in contrast at the downstream of the heated region, a higher Reynolds number results in a lower superheat degree, a slower expansion of the bubble size, and a lower Nusselt number. With increasing the bottom wall thickness, the growth rate of the multiple bubble sizes increases with earlier bubble coalescence because the heat conduction in the solid wall can increase the upstream temperature in the microchannel. The insights gained in this study can be useful as a guide in the design and optimization for a wide range of applications of microchannel thermal management systems.

\section*{Declaration of Competing Interest}
None.

\section*{Acknowledgements}
This work was supported by the National Natural Science Foundation of China (Grant Nos.\ 51920105010, and 51921004).

\section*{Nomenclature}

\begin{tabbing}
$AAAAA$	          \= Surface Area, m$^2$ \kill %this line is just to set tabbing width
$A$				\> Surface Area, m$^2$ \\
$c_p$			\> Specific heat capacity, J/(kg$\cdot$K) \\
$D_h$			\> Hydraulic diameter, m \\
${{\mathbf{f}}_{\sigma }}$	\> Surface tension body force, N/m$^3$ \\
$\mathbf{g}$ 			    \> Gravitational acceleration, m/s$^2$ \\
$h$ 			\> Heat transfer coefficient, W/(m$^2\cdot$K) \\
$h_{lv}$ 		\> Latent heat of vaporization, J/kg \\
$H_b$ 			\> Bottom wall thickness, m \\
$H_c$ 			\> Microchannel height, m \\
$H_t$ 			\> Top wall thickness, m \\
$h_s$ 			\> Specific enthalpy, J/kg \\
$k$		\>Thermal conductivity, W/(m$\cdot$K) \\
$L$ 			\> Length of microchannel, m \\
$\mathbf {n}$ 	\> Normal vector to surface, - \\
$\Nu$           \> Nusselt number, - \\
$p$ 			\> Pressure, Pa \\
$q_\text{int}$ 	\> Solid-fluid interface heat flux, W/m$^2$ \\
${{\dot{q}}_{pc}}$	\> Volumetric phase change source term, W/m$^3$ \\
$q_w$ 			\> Wall heat flux, W/m$^2$ \\
$T$ 			\> Temperature, K \\
$T_\text{int}$ 	\> Solid-fluid interface temperature, K \\
$T_\text{sat}$ 	\> Saturation temperature, K \\
$t$ 			\> Time, s \\
$\mathbf {u}$ 	\> Instantaneous velocity, m/s \\
${{\dot{v}}_{lv}}$ 	\> Mass source term, 1/s \\
$V_b$ 	        \> Bubble volume, m$^3$ \\
$V_{b0}$ 	    \> Initial bubble volume, m$^3$ \\
$V_{L}$ 	    \> Leading bubble volume, m$^3$ \\
$V_{T}$ 	    \> Trailing bubble volume, m$^3$ \\
$V_\text{total}$ 	\> Total bubble volume, m$^3$ \\
$W_b$ 			\> Bottom width, m \\
$W_c$ 			\> Microchannel width, m \\
$W_f$ 			\> Microchannel side wall thickness, m \\
$x$, $y$, $z$ 	\> Coordinate in $x$, $y$, and $z$ directions, m \\
$x_n$ 	\> Position of the bubble nose, m \\
\\
\emph{Greek letters}\\
$\alpha$		\>Volume fraction, -  \\
${{\dot{\alpha }}_{pc}}$ 	\>Source term for phase change, 1/s \\
$\kappa$		\>Interface curvature, 1/m \\
$\mu$			\>Dynamic viscosity, kg/(m$\cdot$s) \\
$\rho$	 		\>Density, kg/m$^3$ \\
$\sigma$		\>Surface tension coefficient, N/m \\
$\varphi $		\>Thermophysical property ($\rho$, $\mu$, $k$) \\
\\
\emph{Superscripts}\\
$l$ 			\>Liquid phase \\
$s$				\>Solid \\
$v$				\>Vapor phase \\
\\
\emph{Subscripts}\\
$\text{bottom}$			\>Bottom wall \\
$\text{int}$			\>Interface \\
$lv$			\>Liquid-vapor phase change \\
$n$			\>Bubble nose \\
$w$ 			\>Wall
\end{tabbing}

\bibliographystyle{elsarticle-num}
\bibliography{BubbleInteraction}

\end{document}